\documentclass[useAMS,usenatbib]{mn2e}
\usepackage{lmacs}
\usepackage{graphicx}
\usepackage[T1]{fontenc} 
\usepackage{aecompl}
\usepackage{aas_macros}
\usepackage{color, soul}
\usepackage{gensymb}
\usepackage{amsmath}
\usepackage{url}
\graphicspath{{./figs/}}
\def\altaffilmark#1{$^{#1}$}
\def\altaffiltext#1#2{$^{#1}$#2}
\newcounter{aaffilcoun}
\newcommand\theaaffil{\addtocounter{aaffilcoun}{1}\theaaffilcoun}
\newcounter{affilcoun}
\newcommand\theaffil{\addtocounter{affilcoun}{1}\theaffilcoun}
\usepackage{xargs}                      

\title[The Highly Luminous Type Ibn Supernova ASASSN-14ms]{The Highly Luminous Type Ibn Supernova ASASSN-14ms}

\author[Vallely et al.]
{P. J. Vallely\altaffilmark{\theaaffil},
J. L. Prieto\altaffilmark{\theaaffil,\theaaffil},
K. Z. Stanek\altaffilmark{1,\theaaffil},
C. S. Kochanek\altaffilmark{1,4},
T. Sukhbold \altaffilmark{1,4},
\newauthor
D. Bersier\altaffilmark{\theaaffil},
J. S. Brown\altaffilmark{1},
Ping Chen\altaffilmark{\theaaffil,\theaaffil},
Subo Dong\altaffilmark{7},
E. Falco\altaffilmark{\theaaffil},
P. Berlind\altaffilmark{8},
M. Calkins\altaffilmark{8},
\newauthor
R. A. Koff\altaffilmark{\theaaffil},
S. Kiyota\altaffilmark{\theaaffil},
J. Brimacombe\altaffilmark{\theaaffil},
B. J. Shappee\altaffilmark{\theaaffil},
T. W.-S. Holoien\altaffilmark{\theaaffil}
\newauthor
T. A. Thompson\altaffilmark{1,4},
M. D. Stritzinger\altaffilmark{\theaaffil}
\\
\altaffiltext{\theaffil}{Department of Astronomy, The Ohio State University, 140 West 18th Avenue, Columbus, OH 43210, USA} \\
\altaffiltext{\theaffil}{N\'ucleo de Astronom\'ia de la Facultad de Ingenier\'ia, Universidad Diego Portales, Av. Ej\'ercito 441, Santiago, Chile} \\
\altaffiltext{\theaffil}{Millennium Institute of Astrophysics, Santiago, Chile} \\
\altaffiltext{\theaffil}{Center for Cosmology and AstroParticle Physics, The Ohio State University, 191 W. Woodruff Ave., Columbus, OH 43210, USA} \\
\altaffiltext{\theaffil}{Astrophysics Research Institute, Liverpool John Moores University, 146 Brownlow Hill, Liverpool L3 5RF, UK} \\
\altaffiltext{\theaffil}{Department of Astronomy, Peking University, Yi He Yuan Road 5, Hai Dian District, Beijing 100871, China} \\
\altaffiltext{\theaffil}{Kavli Institute for Astronomy and Astrophysics, Peking University, Yi He Yuan Road 5, Hai Dian District, Beijing 100871, China} \\
\altaffiltext{\theaffil}{Harvard-Smithsonian Center for Astrophysics, 60 Garden St., Cambridge, MA 02138, USA}\\
\altaffiltext{\theaffil}{Antelope Hills Observatory 980 Antelope DR W Bennett, CO 80102 USA} \\
\altaffiltext{\theaffil}{Variable Stars Observers League in Japan (VSOLJ), 7-1 Kitahatsutomi, Kamagaya 273-0126, Japan} \\
\altaffiltext{\theaffil}{Coral Towers Observatory, Cairns, Queensland 4870, Australia} \\
\altaffiltext{\theaffil}{Institute  for  Astronomy,  University  of  Hawaii,  2680  Woodlawn Drive, Honolulu, Hawaii 96822, USA} \\
\altaffiltext{\theaffil}{Carnegie Observatories, 813 Santa Barbara Street, Pasadena, CA 91101, USA} \\
\altaffiltext{\theaffil}{Department of Physics and Astronomy, Aarhus University, Ny Munkegade 120, DK-8000 Aarhus C, Denmark} \\
}

\begin{document}

\date{Accepted xxx Received xx; in original form xxx}

\pagerange{\pageref{firstpage}--\pageref{lastpage}} \pubyear{2017}

\maketitle

\label{firstpage}

\begin{abstract}
We present photometric and spectroscopic follow-up observations of the highly luminous Type Ibn supernova ASASSN-14ms, which was discovered on UT 2014-12-26.61 at $m_V \sim 16.5$. With a peak absolute $V$-band magnitude brighter than $-20.5$, a peak bolometric luminosity of $1.7 \times 10^{44}$ ergs s$^{-1}$, and a total radiated energy of $2.1 \times 10^{50}$ ergs, ASASSN-14ms is one of the most luminous Type Ibn supernovae yet discovered. In simple models, the most likely power source for this event is a combination of the radioactive decay of $^{56}$Ni and $^{56}$Co at late times and the interaction of supernova ejecta with the progenitor's circumstellar medium at early times, although we cannot rule out the possibility of a magnetar-powered light curve. The presence of a dense circumstellar medium is indicated by the intermediate-width He~I features in the spectra. The faint ($m_g \sim 21.6$) host galaxy SDSS J130408.52+521846.4 has an oxygen abundance below $12+\log(O/H) \lesssim 8.3$, a stellar mass of $M_* \sim 2.6 \times 10^8 M_{\odot}$, and a star formation rate of $\textrm{SFR} \sim 0.02$ $M_{\odot}$ yr$^{-1}$.
\end{abstract}
\begin{keywords}
supernovae: general -- supernovae: individual: ASASSN-14ms -- techniques: photometric -- techniques: spectroscopic
\end{keywords}

\section{Introduction}

A particularly interesting group of supernovae (SNe) are the Type IIn and Type Ibn events. These SNe are characterized by the presence of narrow or intermediate-width hydrogen (IIn) and helium  (Ibn) emission features in their spectra \citep{SNtypespaper,PastorelloIbn}. Their unusual spectra are probably due to interaction between the supernova ejecta and a relatively dense circumstellar medium (CSM) surrounding the progenitor star. As such, they provide valuable insight into the progenitor star by probing its mass-loss rate and composition immediately prior to explosion \citep{TypeIInSchlegel,HosseinzadehIbnMaxLight}.

Although there is some evidence that the historical M~31 supernova SN 1885A was a SN Ibn \citep{PastorelloIbn}, the first clear detection of narrow helium emission lines in the spectra of a hydrogen-deficient (Type Ib/c) supernova was reported for SN 1999cq by \cite{99cqMatheson}. In the ensuing years, a number of additional events showing similar He~I emission features were discovered (SN 2000er and SN 2002ao, \citealt{PastorelloIbn}; SN 2006jc, \citealt{Foley2006jc}), and the small class of objects were termed Type Ibn supernova by \cite{PastorelloIbn}. With our addition of ASASSN-14ms, there are now $\sim 30$ such events known \citep{HosseinzadehIbnMaxLight,2015UShivvers,14ddClassification,2016SmarttPS15dpn,iPTF13beoGorbikov}. There are also two examples of transitional cases that show strong narrow/intermediate-width emission features of both hydrogen and helium   (SN 2005la, \citealt{Pastorello2005la}; SN 2011hw, \citealt{Smith2011hw}). These transitional case events seem to be rare even compared to the small numbers of SNe~Ibn.

The archetypal SNe~Ibn is SN 2006jc, the most well-studied member of this class of transients \citep{Anupama2006jc,HosseinzadehIbnMaxLight,PastorelloIbn,Pastorello06jcOutburst}. \cite{2006CBETNakano} noted that SN 2006jc was spatially coincident with a weaker optical transient which had been observed two years earlier. One potential explanation is the explosion of a Wolf-Rayet star that had either recently undergone a Luminous Blue Variable (LBV) like outburst or had a massive LBV binary companion \citep{Pastorello06jcOutburst}. Such scenarios would be consistent with the idea that the progenitors of Type Ib/c supernovae are massive stars $(M_{ZAMS} \gtrsim 8 M_{\odot})$. 

Among SNe~Ibn, \cite{HosseinzadehIbnMaxLight} further distinguish between the ``P Cygni'' and ``emission'' subclasses based on differences in the early spectra. Events in the P Cygni subclass are characterized by the presence of narrow P Cygni profiles in their spectra, while those in the emission subclass are dominated by broader emission lines. Virtually all known SNe~Ibn fall into one of these two subclasses, with approximately half of the total population falling into each classification. It is currently unclear whether these distinctions trace two different CSM configurations or a continuum of CSM properties.

Here, we describe the discovery and follow-up observations of the highly luminous Type Ibn supernova ASASSN-14ms and present simple physical models for its unusual light curve. In Section~\ref{sec:obs} we describe the discovery of this transient by ASAS-SN, report the results of our photometric and spectroscopic follow-up observations, and also briefly discuss the spectroscopic characteristics of its faint host galaxy, SDSS J130408.52+521846.4 at $z=0.0540$. In Section~\ref{sec:comparison} we compare the photometric and spectroscopic observations of ASASSN-14ms to those of other SNe~Ibn. In Section~\ref{sec:LCModels} we briefly describe a few simple physical light curve models and then use these models to fit the observed light curve of ASASSN-14ms. In Section~\ref{sec:conclusion} we discuss our conclusions.

\section{Observations and Analysis}
\label{sec:obs}

\subsection{Discovery}
\label{subsec:disc}

ASASSN-14ms was discovered by the All-Sky Automated Survey for Supernovae \citep[ASAS-SN;][]{ShappeeASASSN,ASASSNCatalog} in images obtained on JD 2457018.11 (UT 2014-12-26.61) at J2000 R.A. 13$^\textnormal{h}$04$^\textnormal{m} 08\fs69$ and Decl. +52{\degree}18'$46\farcs5$. These images were obtained using the quadruple 14cm ASAS-SN telescope Brutus, which is deployed at the Haleakala station of the Las Cumbres Observatory Global Telescope Network \citep[LCOGT;][]{Brown2013}. Its discovery was confirmed two days later by amateur astronomer S. Kiyota using a Planewave CDK 0.61m telescope located at the Sierra Remote Observatory \citep{14msDiscoveryATel}. Photometry obtained on JD 2457018.11 gave an apparent $V$-band magnitude of 16.5 mag, and the transient was not detected in images obtained on JD 2457007.10 (UT 2014-12-15.60) and earlier down to a limiting magnitude of $V \sim 17.5$ mag. We find no evidence in archival ASAS-SN photometry for a pre-supernova outburst analogous to that observed for SN 2006jc. However, with ASASSN-14ms peaking at $V \sim 16.5$ and a survey depth of $V \sim 17.5$, such an outburst would have been too faint to be detected.

ASASSN-14ms is located $\sim 1\farcs1$ South and $\sim 0\farcs8$ East of the center of SDSS J130408.52+521846.4. Prior to this work, no spectra of this galaxy were publicly available. A follow-up spectrum obtained using the Multi-Object Double Spectrographs \citep[MODS;][]{PoggeMODS} on the Large Binocular Telescope indicates that the host galaxy has a redshift of $z = 0.0540 \pm 0.0001$. Adopting $H_0=69.6$ and $\Omega_M=0.286$ and assuming a flat universe yields a luminosity distance of 242 Mpc \citep{2006Wright}. The host galaxy spectrum is discussed in greater detail in Subsection~\ref{subsec:hostspec}.

\subsection{Photometry}
\label{subsec:phot}

Photometric observations were obtained over the course of 110 days following the discovery of ASASSN-14ms. The supernova remained bright enough to be detected in ASAS-SN images for nearly a month after discovery. ASAS-SN images are processed in an automated pipeline using the ISIS image subtraction package \citep{1998Alard,2000Alard}. Using the IRAF \textsc{apphot} package, we performed aperture photometry on the subtracted images and then calibrated the results using the AAVSO Photometric All-Sky Survey \citep[APAS;][]{Henden2015}. 

We obtained $g$-band images using the 1m Las Cumbres Observatory telescope deployed at McDonald Observatory, and $V$-band images from a 0.25m telescope at the Antelope Hills Observatory operated by R.A. Koff. Most of the post-maximum photometry was obtained using the 2m Liverpool Telescope \citep[LT;][]{Steele2004lT} at the Observatorio del Roque de los Muchachos, which provided \textit{griz} observations spanning nearly 80 days. All images were reduced after bias/dark-frame and flat-field corrections. We extract the photometry using the IRAF \textsc{apphot} package, and the photometric calibrations are performed using APASS for $V$-band and the first data release of Panoramic Survey Telescope and Rapid Response System \citep[Pan-STARRS;][]{PanSTARRS2016Flewelling} for SDSS $g$, $r$, $i$ and $z$ bands. In order to correct for the host galaxy contribution, we performed image subtraction on the LT observations using images of the host obtained long after the supernova faded (April 2017). We note that for the last three epochs of photometric observations the supernova is several magnitudes dimmer than the host, making the photometry somewhat suspect.

\begin{figure*}
\centering
\includegraphics[scale=0.41]{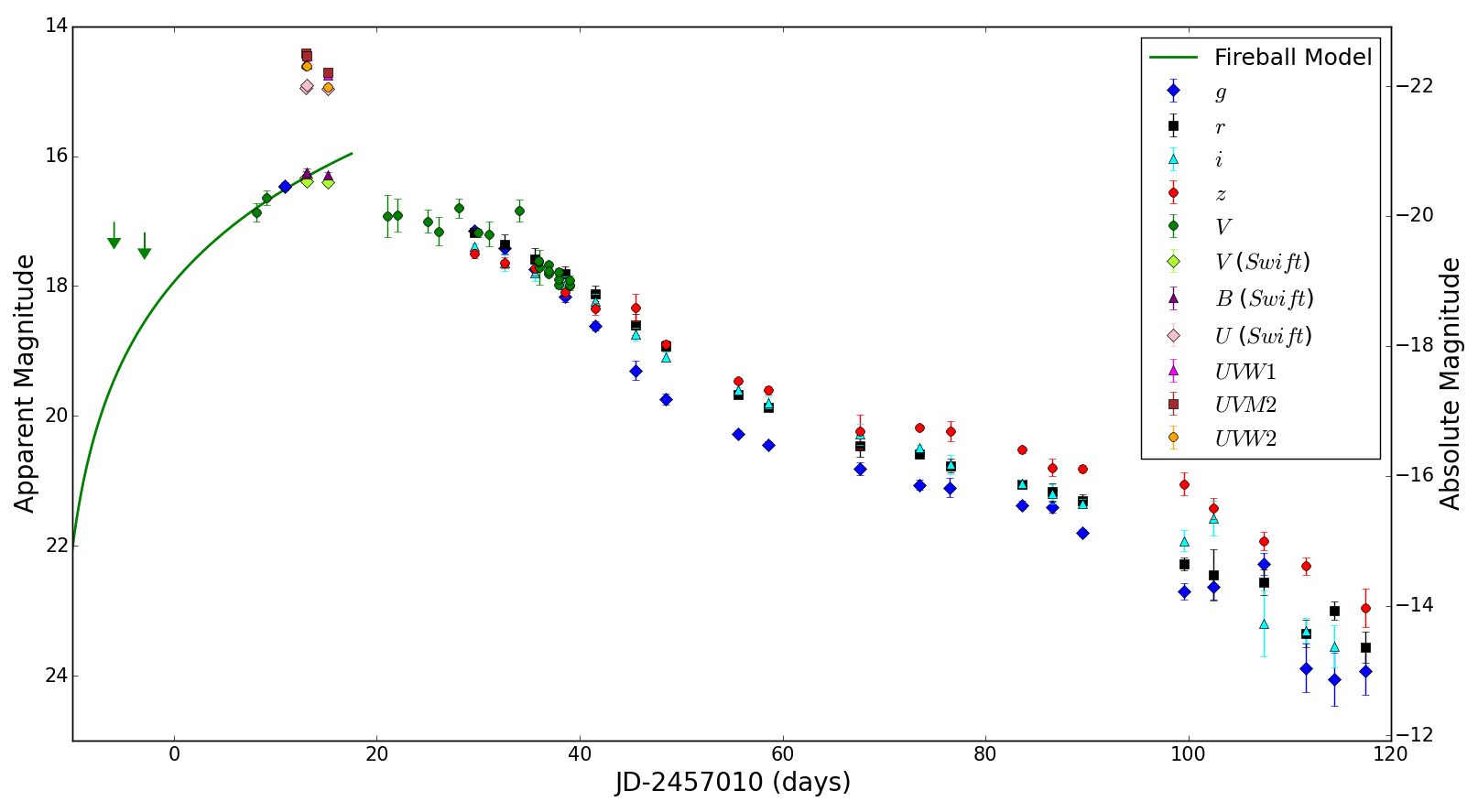}
\caption{The complete photometry of ASASSN-14ms with the marker shapes and colors indicating the filter. Error bars are shown for all points, but can be smaller than the symbol used to represent the data. The two arrows show $3 \sigma$ $V$-band non-detection upper limits from ASAS-SN, and the green curve shows the rising-phase expanding fireball fit discussed briefly in Section~\ref{subsec:phot}.}
\label{fig:LightCurves}
\end{figure*}

We also obtained images in the $U$, $B$, $V$, $UVW1$, $UVW2$ and $UVM2$ bands with \textit{Swift}'s Ultraviolet Optical Telescope \citep[UVOT;][]{2005RomingUVOT}. The \textit{Swift}/UVOT photometry is extracted using a $5\farcs0$ aperture and a sky annulus with an inner radius of $15\farcs0$ and an outer radius of $30\farcs0$ with the \textsc{uvotsource} task in the \textsc{heasoft} package. The \textit{Swift}/UVOT photometry is calibrated in the Vega magnitude system based on \cite{2011Breeveld}. We note that typical offsets between the \textit{Swift} $V$-band and Johnson $V$-band photometry are small \citep[$\sim 0.05$ mag;][]{Godoy2017}, and we add a conservative estimate of the uncertainty introduced by the offset of $0.075$ mag in quadrature to the measured photometric uncertainty. The ground-based photometry obtained for ASASSN-14ms is reported in Table~\ref{tab:groundphot}, and the \textit{Swift} photometry is reported in Table~\ref{tab:Swiftphot}. The complete set of photometry is presented in  Figure~\ref{fig:LightCurves}.

With a peak absolute $V$-band magnitude of $-20.5 \pm 0.1$ mag, ASASSN-14ms is consistent with being the most luminous SNe~Ibn yet discovered. The only other events to attain comparably high peak absolute $V$-band magnitudes are  SN 2015U (PSN J07285387+3349106) and iPTF15ul, at approximately $-20.3$ mag \citep{PastorelloVIII,TsvetkovPSN} and $-20.8$ mag \citep{HosseinzadehIbnMaxLight}, respectively. Within published uncertainties, the peak $V$-band magnitudes of all three events are consistent with one another, and significant uncertainties associated with estimating the host galaxy extinction for SN~2015U and iPTF15ul make it difficult to confidently determine which event is truly the most luminous at peak.

ASASSN-14ms was detected relatively close to maximum light. As a result, the rise of the light curve is poorly characterized. Nevertheless, we can fit a simple expanding fireball model to this limited data to estimate the explosion date. The key assumption of the expanding fireball model is that the luminosity of the supernova should scale as $f={\alpha}{\times}(t-t_0)^2$ \citep{Riess1999,Nugent2011}. We fit this functional form to the $V$-band fluxes and obtained an estimated explosion date of JD $2456998.0$ (UT 2014-12-06.5). However, since this fit uses only a handful of points near the end of the rising phase it is unclear how accurate this value is.

To estimate the date of maximum light, we calculated an uncertainty-weighted fit to the combined set of near-peak $V$-band observations using a third degree polynomial. This interpolation yields JD $2457025.7 \pm 0.5$ (UT 2015-01-03) for the epoch of maximum light and a maximum apparent $V$-band magnitude of 16.4. We will reference our spectroscopic observations to this date.

We estimate bolometric luminosities using a Markov Chain Monte Carlo (MCMC) routine to fit a blackbody curve to the observed spectral energy distribution. We first fit the dates with sufficient filter coverage (i.e., dates with \textit{Swift} and LT observations) to produce well-constrained temperature estimates. We then built a temperature prior by fitting a fifth degree polynomial to these constrained temperature estimates. For photometry obtained within the range of the \textit{Swift} and LT observations we simply take this polynomial interpolation to be the temperature prior. For the handful of pre-maximum $V$-band observations without multi-filter coverage we use a constant temperature prior equal to the 17,300 K value determined for the earliest date \textit{Swift} observations in order to avoid extrapolation. The temperature prior is applied when fitting dates for which we have observations in less than four filters.

\begin{figure}
\centering
\includegraphics[scale=0.41]{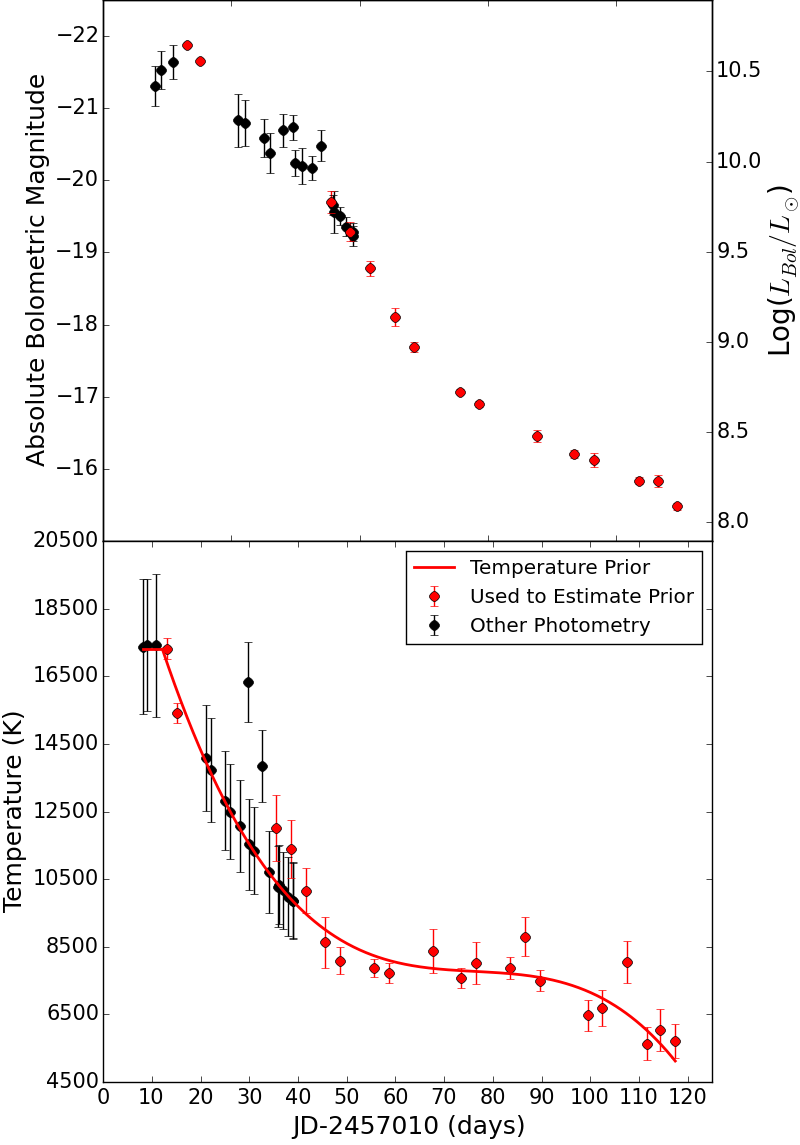}
\caption{The bolometric light curve (top) and temperature evolution (bottom) of ASASSN-14ms. Errors are shown for all points, but can be smaller than the symbol used to represent the data. Red symbols indicate data used to build the temperature prior. The temperature prior used to calculate the bolometric light curve is shown in the lower panel. The bolometric light curve (including luminosity, temperature, and radius values) is available in the online version of the paper.}
\label{fig:BolometricLC}
\end{figure}

All photometry was corrected for a Galactic foreground extinction of $E(B-V)=0.01$ mag \citep{Schlafly11} before being fit. The UV colors of the transient are relatively blue and we see no evidence of Na~ID absorption in the spectra, so we make no additional correction for extinction in the host galaxy. The resulting bolometric light curve is shown with $1\sigma$ error estimates in Figure~\ref{fig:BolometricLC}. Uncertainties are shown for all points, although for many of the \textit{Swift} and LT dates the symbols used to plot the data are larger than the uncertainties. The bolometric light curve of ASASSN-14ms is compared to those of other SNe~Ibn in Section~\ref{sec:comparison}.

We can integrate the bolometric luminosity over time in the supernova rest frame to estimate that the total radiated energy is $(2.1 \pm 0.3) \times 10^{50}$ ergs. This is a lower limit since it does not include emission outside the time span of our observations. In Section~\ref{sec:LCModels} we discuss refinements to this estimate based on our semi-analytic physical models.

\begin{table*}
\caption{Ground-Based Photometric Observations.}
\label{tab:groundphot}
\begin{tabular}{ccccccc}
\hline\hline
JD & Telescope & $V$ & $g$ & $r$ & $i$ & $z$ \\
\hline\hline 
2457018.1 & ASAS-SN: 14cm & $16.86 \pm 0.14$ & -- & -- & -- & -- \\
2457019.1 & ASAS-SN: 14cm & $16.64 \pm 0.11$ & -- & -- & -- & -- \\
2457020.906 & LCOGT: 1m & --  & $16.471 \pm 0.059$ & -- & -- & -- \\
2457020.911 & LCOGT: 1m & --  & $16.460 \pm 0.059$ & -- & -- & -- \\
2457031.1 & ASAS-SN: 14cm & $16.92 \pm 0.32$ & -- & -- & -- & -- \\
2457032.1 & ASAS-SN: 14cm & $16.91 \pm 0.25$ & -- & -- & -- & -- \\
2457035.1 & ASAS-SN: 14cm & $17.00 \pm 0.18$ & -- & -- & -- & -- \\
2457036.1 & ASAS-SN: 14cm & $17.16 \pm 0.22$ & -- & -- & -- & -- \\
2457038.1 & ASAS-SN: 14cm & $16.80 \pm 0.15$ & -- & -- & -- & -- \\
2457039.6 & LT: 2m & --  & $17.15 \pm 0.063$ & $17.17 \pm 0.066$ & $17.39 \pm 0.053$ & $17.50 \pm 0.065$ \\
2457040.0 & AHO: 25cm & $17.181 \pm 0.065$ & -- & -- & -- & -- \\
2457041.0 & ASAS-SN: 14cm & $17.20 \pm 0.19$ & -- & -- & -- & -- \\
2457042.6 & LT: 2m & --  & $17.42 \pm 0.083$ & $17.35 \pm 0.16$ & $17.64 \pm 0.13$ & $17.64 \pm 0.085$ \\
2457044.0 & ASAS-SN: 14cm & $16.84 \pm 0.17$ & -- & -- & -- & -- \\
2457045.6 & LT: 2m & --  & $17.73 \pm 0.098$ & $17.58 \pm 0.16$ & $17.79 \pm 0.13$ & $17.72 \pm 0.11$ \\
\vdots & \vdots & \vdots & \vdots & \vdots & \vdots & \vdots \\
\hline\hline
\end{tabular} \\
\begin{flushleft}
This table is available in its entirety in machine-readable format in the online version of the paper.
\end{flushleft}
\end{table*}

\begin{table*}
\caption{\textit{Swift} UVOT Photometric Observations}
\label{tab:Swiftphot}
\begin{tabular}{ccccccc}
\hline\hline
JD & $V$ & $B$ & $U$ & $UVW1$ & $UVM2$ & $UVW2$ \\
\hline\hline
2457023.0 & $16.328 \pm 0.078$ & $16.273 \pm 0.048$ & $14.943 \pm 0.042$ & $14.561 \pm 0.045$ & $14.418 \pm 0.052$ & $14.606 \pm 0.042$ \\
2457023.1 & $16.380 \pm 0.079$ & $16.240 \pm 0.047$ & $14.906 \pm 0.042$ & $14.586 \pm 0.045$ & $14.457 \pm 0.043$ & $14.606 \pm 0.042$ \\
2457025.1 & $16.396 \pm 0.055$ & $16.289 \pm 0.039$ & $14.956 \pm 0.037$ & $14.751 \pm 0.042$ & $14.708 \pm 0.045$ & $14.935 \pm 0.040$ \\
\hline\hline
\end{tabular} \\
\begin{flushleft}
This table is available in machine-readable format in the online version of the paper.
\end{flushleft}
\end{table*}

\subsection{Spectroscopy}
\label{subsec:spec}

We obtained six optical spectra of ASASSN-14ms, all after maximum light, along with an optical spectrum of the host galaxy SDSS J130408.52+521846.4. The first three spectra were obtained at phases of +7 days, +11 days, and +12 days relative to the $V$-band peak using the Fast Spectrograph on the (1.5m) Tillinghast Telescope located at the Fred Lawrence Whipple Observatory \citep[FAST;][]{FabricantFAST,1997Tokarz}. Two spectra were obtained using the Multi-Object Double Spectrographs mounted on the 8.4m Large Binocular Telescope \citep[MODS;][]{PoggeMODS} at phases of +18 and +44 days. The sixth spectrum was obtained at a phase of +39 days using The Ohio State Multi-Object Spectrograph mounted on the 2.4m Hiltner Telescope at the MDM Observatory \citep[OSMOS;][]{MartiniOSMOS}. Finally, a spectrum of the host galaxy was obtained using MODS long after the supernova had faded. This spectrum is discussed further in Section~\ref{subsec:hostspec}. These spectroscopic observations are summarized in Table~\ref{tab:spectra}.

The MODS spectra (including that of the host galaxy) were reduced with a combination of the \textsc{modsccdred}\footnote{\url{http://www.astronomy.ohio-state.edu/MODS/Software/modsCCDRed/}} \textsc{python} package and the \textsc{modsidl} pipeline\footnote{\url{http://www.astronomy.ohio-state.edu/MODS/Software/modsIDL/}}. We performed aperture photometry on the $r'$-band acquistion image to place the spectroscopic flux measurements on an absolute scale. In order to reduce the FLWO and MDM spectra we used standard techniques in IRAF to reduce the 2D images and to extract and calibrate the 1D spectra in wavelength and flux. The Supernova Identification code \citep[SNID;][]{TonrySNIDAlgorithm,BlondinSNID}, classifies all six of the ASASSN-14ms spectra as either Type Ib or Type Ic supernovae. Since SNID does not include the Type Ibn sub-classification this result is consistent with a SN Ibn classification because, aside from their helium emission features, SNe Ibn spectra broadly resemble those of SNe Ib/c.

\begin{figure}
\centering
\includegraphics[scale=0.42]{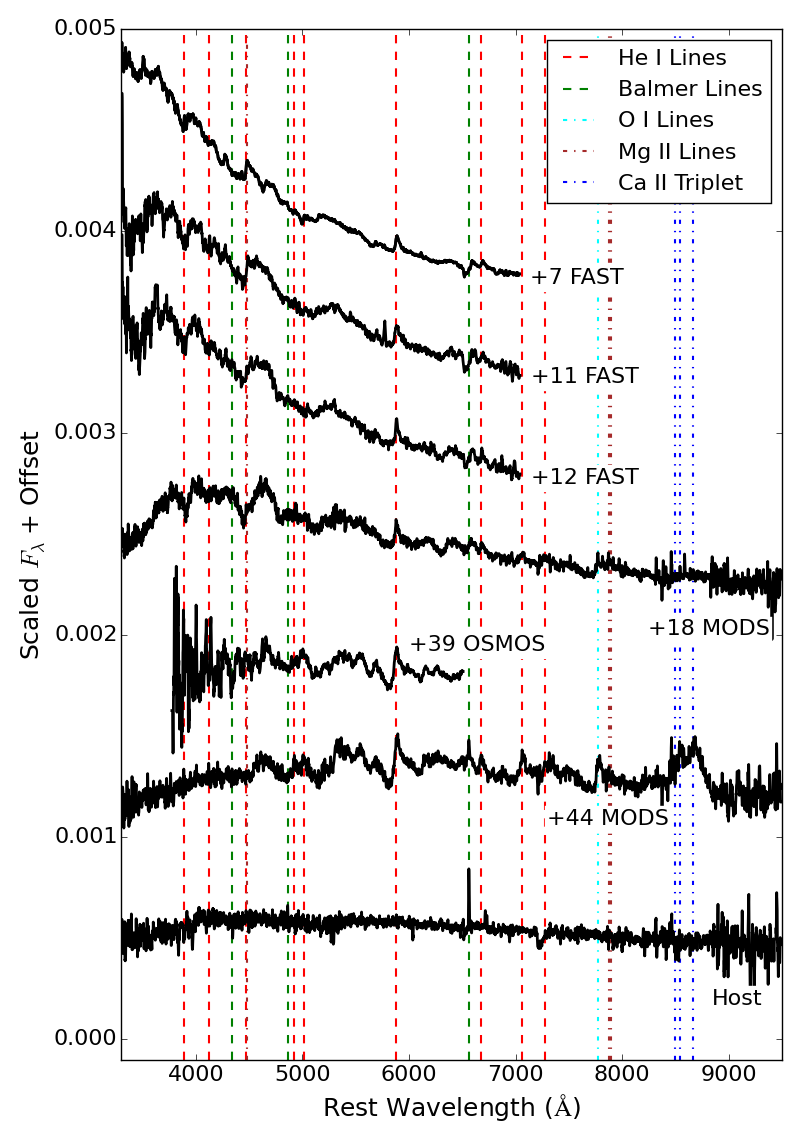}
\caption{Rest-frame spectra of ASASSN-14ms and the spectrum of SDSS J130408.52+521846.4. No correction is applied for extinction. The locations of various atomic transitions are shown using the vertical dashed lines, and the spectra are labeled according to their phase relative to the $V$-band peak. The spectra have been modestly smoothed. Note the consistent presence of He~I emission features throughout the spectra.}
\label{fig:LineIDs}
\end{figure}

The six spectra are shown in Figure~\ref{fig:LineIDs} along with colored vertical lines marking the location of relevant O~I, Mg~II, Ca~II, He~I and H~Balmer lines. There is strong evidence for the presence of He~I lines (the red vertical lines in Figure~\ref{fig:LineIDs}) in all of the spectra. We see five He~I lines in the +7, +11, and +12 days FAST spectra (at 3889~{\AA}, 4471~{\AA}, 5016~{\AA}, 5876~{\AA}, and 6678~{\AA}). In the noisier +39 days OSMOS spectrum we see only the He~I 5876~{\AA} feature clearly. In the +18 days and +44 days MODS spectra we again see the 4922~{\AA}, 5876~{\AA}, 6678~{\AA}, and 7065~{\AA} He~I lines, but the 3889~{\AA} feature can no longer be clearly detected. The FWHM of the He~I emission features all lie within the range $\sim$1000 - 2000 km/s, which is consistent with the narrow/intermediate width He~I features identified in other SNe~Ibn. Only weak H$\alpha$ signatures are seen in the spectra. This is consistent with the spectra of other SNe~Ibn, as we discuss in Section~\ref{sec:comparison}.

The greater wavelength coverage of the two MODS spectra allows us to identify several ions not detected in the FAST and OSMOS spectra, namely Ca~II, O~I, and Mg~II. The strength of the emission lines for these three ions is stronger in the +44 days spectrum than in the +18 days spectrum. In both MODS spectra we see strong emission from the near-IR Ca~II triplet, as well as from the O~I 7774~{\AA} triplet. There are also emission signatures from the Mg~II 7877~{\AA} and 7896~{\AA} lines, although these are not as distinct. Confident identification of Mg~II is complicated by the presence of strong He~I 4471~{\AA} features, which mask the presence of the Mg~II 4481~{\AA} features. Ca~II, O~I, and Mg~II features were also identified in the spectra of SN 2006jc \citep{Foley2006jc,Anupama2006jc}, making their detection here unsurprising.

The inferred photospheric expansion velocity, estimated by measuring the absorption minimum of P-Cygni profiles in the spectra, increases with time. From several of the strong He~I features (4471~{\AA}, 5016~{\AA}, 5876~{\AA}, and 6678~{\AA}) we find that the three FAST spectra (phases +7, +11, and +12 days) exhibit comparable expansion velocities of approximately 1,200-1,600 km/s. In the phase +18 days spectrum, we estimate the expansion velocity from the He~I 4471~{\AA} and 5876~{\AA} features and find that it has increased to 2,000-2,300 km/s. For the +39 days OSMOS spectrum we are limited to measuring the He~I 5876~{\AA} feature, but the 3,700-4,000 km/s velocity estimate we obtain is consistent with that measured for the +44 days MODS spectrum using the O~I 7774~{\AA}, Ca~II 8498~{\AA}, and He~I 5876~{\AA} features. This increasing expansion velocity is likely the result of the supernova ejecta shocking and accelerating the surrounding CSM.

\begin{table*}
\caption{ASASSN-14ms Spectroscopic Observations.}
\label{tab:spectra}
\begin{tabular}{ccccccc}
\hline\hline
JD & Phase (Days) & Telescope & Instrument & Exposure (s) & Wavelength Coverage & Resolution\\
\hline\hline
2457033.0 & +7 & FLWO-1.5m & FAST & 3600 & 3480 {\AA} -- 7420 {\AA} & $\sim$3 {\AA} \\
2457037.0 & +11 & FLWO-1.5m & FAST & 1800 & 3480 {\AA} -- 7420 {\AA} & $\sim$3 {\AA} \\ 
2457038.0 & +12 & FLWO-1.5m & FAST & 1800 & 3480 {\AA} -- 7420 {\AA} & $\sim$3 {\AA} \\ 
2457043.9 & +18 & LBT-8.4m & MODS & 1800 & 3400 {\AA} -- 10000 {\AA} & $\sim$3 {\AA} \\
2457065.0 & +39 & MDM-2.4m & OSMOS & 4800 & 3980 {\AA} -- 6860 {\AA} & $\sim$4 {\AA} \\
2457069.8 & +44 & LBT-8.4m & MODS & 2700 & 3400 {\AA} -- 10000 {\AA} & $\sim$3 {\AA} \\
2457901.7$^\textnormal{a}$ & +876 & LBT-8.4m & MODS & 7200 & 3000 {\AA} -- 9500 {\AA} & $\sim$3 {\AA} \\
\hline\hline
\end{tabular} \\
$^\textnormal{a}$ Spectrum of the host galaxy, obtained long after ASASSN-14ms had faded.
\begin{flushleft}
Note that phase is calculated relative to the inferred maximum light date of JD 2457025.7.
\end{flushleft}
\end{table*}

\subsection{Host Galaxy SDSS J130408.52+521846.4}
\label{subsec:hostspec}

In order to constrain the properties of the host galaxy SDSS J130408.52+521846.4, we performed a search of archival databases.We retrieved the archival $ugriz$ model magnitudes from the Twelfth Data Release of the Sloan Digital Sky Survey \citep[SDSS;][]{Alam15}. The host was also detected in the NUV filter by the Galaxy Evolution Explorer \citep[GALEX;][]{Martin05}. The host is not significantly spatially extended on the sky, so there is no worry that the cataloged magnitudes sample drastically different spatial scales. The host is too faint to be detected in the 2MASS or WISE surveys, and there are no archival Spitzer, Herschel, HST, Chandra, or X-ray Multi-Mirror Mission (XMM-Newton) observations of the host galaxy. There are also no associated radio sources in the FIRST \citep{Becker95} or NVSS \citep{Condon98} catalogs.

We modeled the SDSS+GALEX host galaxy SED with the Fitting and Assessment of Synthetic Templates \citep[FAST;][]{Kriek09} code. We adopted a \citet{Cardelli89} extinction law with $R_V = 3.1$ and a Galactic extinction of $A_V = 0.031$ mag \citep{Schlafly11}, an exponentially declining star-formation history, a Salpeter IMF, and the \citet{Bruzual03} stellar population models. We obtained a reasonable fit ($\chi^2_{\nu} = 0.42$) which yielded estimates of the stellar mass and star formation rates of $M_* = (2.6_{-1.5}^{+0.8})\times 10^8 M_{\odot}$ and SFR$\sim 0.02$ $M_{\odot}$ yr$^{-1}$.

We also obtained a late-time LBT MODS spectrum long after the transient had faded (at phase +876 days) in order to assess the spectroscopic properties of the host. The spectrum is continuum dominated, and shows only weak nebular emission features. However, we are able to cleanly detect the H$\alpha$ emission line (EW$\sim14.8${\AA}). We determine the redshift of the host to be $z = 0.0540 \pm 0.0001$, corresponding to a luminosity distance of $242$~Mpc. The H$\alpha$ luminosity ($L_{\rm H\alpha}$ $\sim 3.3\times10^5$ L$_{\odot}$) corresponds to a star formation rate of $\sim9.8\times10^{-3}$~M$_{\odot}$~yr$^{-1}$ \citep{Kennicutt98}. There is a marginal detection of the N~II~$\lambda 6583$ line, which we use to place an approximate upper limit on the metalliticy of the host galaxy. The line ratio is constrained to be N~II~$\lambda 6583$/H$\alpha$ $\lesssim 0.075$, corresponding to an upper limit on the oxygen abundance of $12+\log(O/H) \lesssim 8.3$ \citep{Pettini04}. This places SDSS J130408.52+521846.4 near the lower end of the $12+\log(O/H)$ dex distribution for SNe~Ibn host galaxies \citep{PastorelloVI,TaddiaMetallicity}.

\section{Comparison to other Type Ibn Supernovae}
\label{sec:comparison}

\begin{figure}
\centering
\includegraphics[scale=0.42]{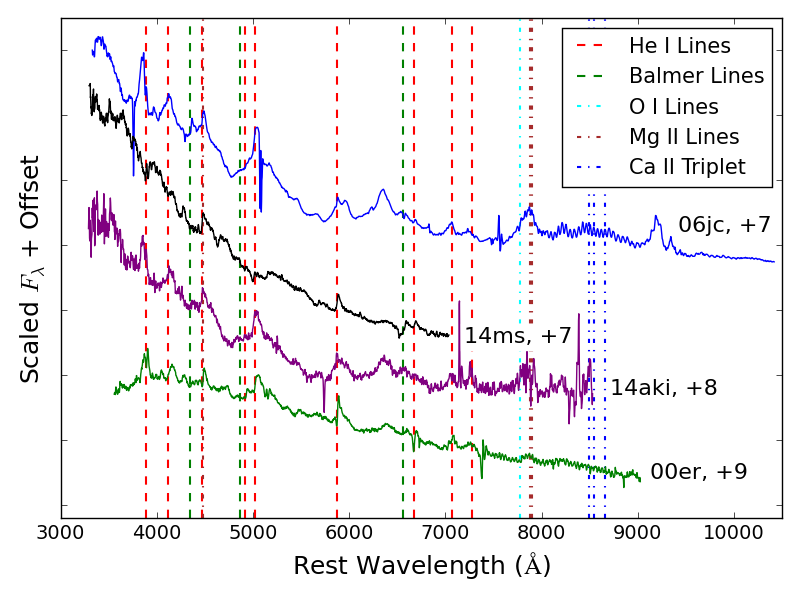}
\includegraphics[scale=0.42]{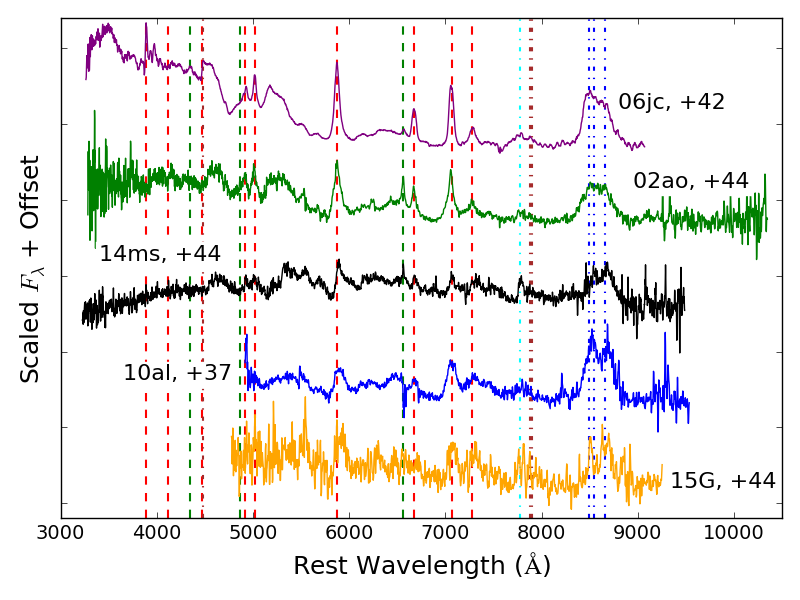}
\caption{The upper panel shows an early (+7 days) spectrum of ASASSN-14ms compared to other Type Ibn supernovae at a similarly early phase. Note that phase is defined relative to the observed date of maximum light. The lower panel shows a late (+44 days) spectrum of ASASSN-14ms as compared to other Type Ibn supernovae at similar phases. Spectra for all supernovae are corrected for Galactic extinction.}
\label{fig:IbnComparisons}
\end{figure}

We compare spectra of ASASSN-14ms to those of other SNe~Ibn in Figure~\ref{fig:IbnComparisons}. Here the spectra are corrected for Galactic extinction using the extinction coefficient data from the \cite{Schlafly11} recalibration of the \cite{1998Schlegel} dust map. The top panel shows early phase spectra obtained approximately one week after maximum light, and the bottom panel shows later phase spectra obtained roughly forty days after maximum light. The three other SNe~Ibn presented in the early phase comparison panel are SN 2006jc \citep{Anupama2006jc}, iPTF14aki \citep{HosseinzadehIbnMaxLight}, and SN 2000er \citep{PastorelloIbn}. The late phase comparison panel adds SN 2006jc \citep{Anupama2006jc}, SN 2002ao \citep{Foley2006jc}, SN 2010al \citep{PastorelloIV}, and SN 2015G \citep{HosseinzadehIbnMaxLight}.

There is a good deal of spectral diversity in the Ibn class at early times \citep{HosseinzadehIbnMaxLight}, and ASASSN-14ms does not seem to be an exception. The presence of He~I P Cygni profiles in its spectrum and the $\sim$1300 km/s FWHM of their emission components places ASASSN-14ms in the P Cygni subclass of SNe~Ibn, like SN 2000er, but its continuum shape seems to more closely resemble that of SN 2006jc and iPTF14aki, both of which are in the emission subclass. At late times, however, ASASSN-14ms generally resembles other SNe~Ibn. At this phase the P Cygni absorption components of events in the P Cygni subclass have weakened, and the continuum slopes have all flattened considerably. Strong emission from the Ca II triplet (the blue vertical lines in Figure~\ref{fig:IbnComparisons}) as well as moderate strength O~I and Mg~II emission features (the cyan and brown vertical lines in Figure~\ref{fig:IbnComparisons}, respectively) are also general features of late-time SNe~Ibn spectra, and ASASSN-14ms exhibits them as well. The similarity of ASASSN-14ms to SN 2006jc and SN2002ao at this phase is particularly strong, including the development of a narrow H$\alpha$ emission line. Thus, from a spectroscopic standpoint, ASASSN-14ms is not a particularly striking outlier among SNe~Ibn. It is a little unusual in the early phases when the class is broadly spectroscopically heterogeneous, and it is fairly normal in the later phases when the class becomes broadly spectroscopically homogeneous.

\begin{figure}
\centering
\includegraphics[scale=0.41]{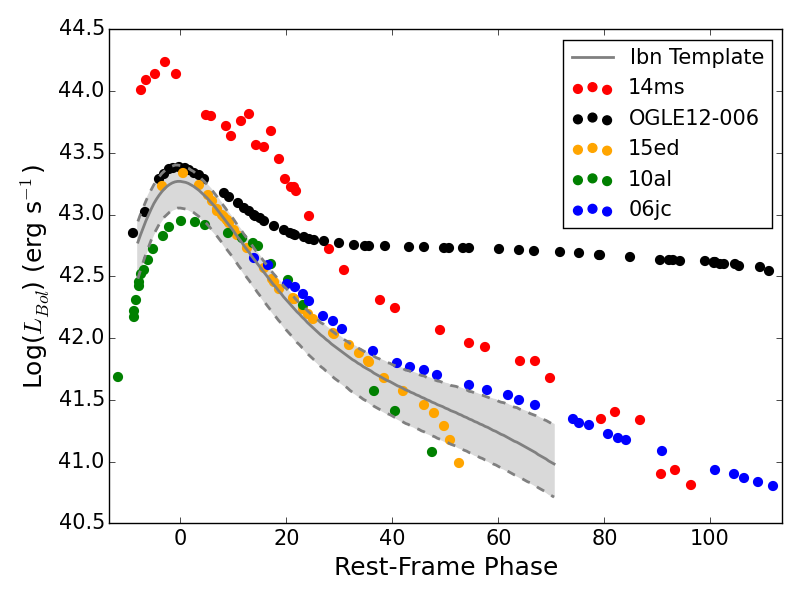}
\caption{The rest-frame bolometric light curves of several Type Ibn supernovae along with the Type Ibn template of Hosseinzadeh et al. (2017). Note the unusually high luminosity of ASASSN-14ms.}
\label{fig:IbnBolometricLCs}
\end{figure}

Its light curve, however, is notably different. The bolometric light curve of ASASSN-14ms is shown in Figure~\ref{fig:IbnBolometricLCs} as compared to a representative sample of SNe~Ibn with publicly available bolometric light curves. The four other individual events are OGLE-2012-SN-006 \citep{PastorelloV}, ASASSN-15ed \citep{PastorelloVII}, SN 2010al \citep{Moriya2010al}, and SN 2006jc \citep{PastorelloIbn}. We also include the Type Ibn light curve template from \cite{HosseinzadehIbnMaxLight}. The range of this template encompasses $\sim 95 \%$ of the photometric points used in their sample of 22 SNe~Ibn and provides a good representation of typical SNe~Ibn bolometric light curve properties.

The ASASSN-14ms light curve is notably broader than other SNe~Ibn. This is indicative of a longer diffusion timescale and a considerably higher total radiated energy. Unlike OGLE-2012-SN-006, another unusually luminous member of this class, ASASSN-14ms does not exhibit an abnormally slow late-time decline and decays at a rate typical of other SNe~Ibn. What makes ASASSN-14ms particularly unusual within the SNe~Ibn class is its extreme luminosity. 

The stark contrast in luminosity is clearly illustrated in Figure~\ref{fig:IbnBolometricLCs}. OGLE-2012-SN-006 and ASASSN-15ed are considered to be examples of unusually luminous SNe~Ibn, and yet the peak luminosities for both lie within the upper limit of the Ibn template of \cite{HosseinzadehIbnMaxLight}. The peak luminosity of ASASSN-14ms, on the other hand, is a factor of $\sim 6$ above this upper limit, and the light curve remains a factor of $\sim 3$ brighter than the template upper limit even at 70 days after maximum light. In terms of luminosity, ASASSN-14ms is a clear outlier among SNe~Ibn.

A likely explanation is that ASASSN-14ms is hotter than other SNe~Ibn. At comparable phases ($\sim 12$ days), SN 2006jc and SN 2010al have $U-B$ color indices of $-0.7$ and $-0.8$, respectively \citep{Pastorello2005la,Moriya2010al}. Although we do not have \textit{Swift} observations of that epoch for ASASSN-14ms, making the comparison imperfect, its near-maximum $U-B$ color index is $-1.3$, indicating that it is considerably hotter than these events. Furthermore, we have \textit{g}- and \textit{r}-band observations for ASASSN-14ms obtained at phases ($\sim 15$ days) comparable to those obtained for ASASSN-15ed by \citet{PastorelloVII}. At this epoch, ASASSN-14ms and ASASSN-15ed have $g-r$ color indices of $-0.03$ and $0.3$, respectively, which is also consistent with ASASSN-14ms being hotter than other SNe~Ibn.

\section{Physical Light Curve Models}
\label{sec:LCModels}

Semi-analytic models have proven to be a useful tool for analyzing supernova light curves. In this section we compare several of these physically motivated models to ASASSN-14ms in order to probe the physics of the explosion and better understand its unusual light curve. These relatively simple models allow us to make reasonable estimates for the explosion parameters in a variety of potential light curve powering scenarios, and provide a basis for estimating the likelihood of each scenario. We consider three distinct powering sources: the radioactive decay of $^{56}$Ni produced during the initial explosion, the deposition of energy into the ejecta by magnetar spin down, and the interaction of ejecta with circumstellar material. We also consider model light curves that implement combinations of these power sources. All models are fit in the rest frame of the supernova, and we exclude the final three epochs of data from our analysis because the supernova is only marginally detected in the host-subtracted photometry.

We present the functional forms of the semi-analytic models we use in the following subsections. These brief descriptions are not intended to constitute a thorough review of light curve modeling techniques, nor are they intended to constitute derivations of the models. Rather, we present these forms to facilitate replication of our results and the implementation of similar analyses in the future. For an in-depth review of the derivations and physical assumptions of these models please see the citations therein.

We also emphasize that we do not perform a rigorous exploration of the entire parameter space associated with these models, nor do we explore the entire range of potential power sources or power source combinations. For instance, we consider only the combination of CSM interaction and radioactive decay. In principle, any combination of these power sources could contribute to the observed light curve. The aim is only to explore physically plausible scenarios to identify parameter combinations that can explain the unusual light curve of ASASSN-14ms.

\begin{figure}
\centering
\includegraphics[scale=0.41]{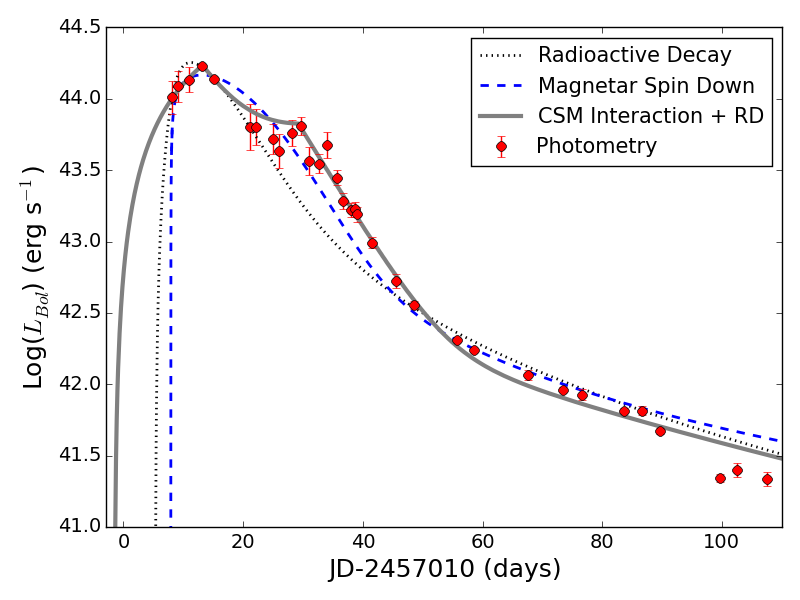}
\caption{The bolometric light curve of ASASSN-14ms compared with several physical models. The gray curve is a model powered solely by the radioactive decay of $^{56}$Ni and $^{56}$Co. The blue curve is a model powered solely by the spin down of a magnetar. The gray curve is a model powered by a combination of CSM-interaction, which dominates at early times, and radioactive decay, which dominates at late times. Note that the radioactive decay model requires the unphysical condition that $M_{Ni}>M_{ej}$. The other two models utilize physically plausible parameters. Note that, while all models are fit in the rest frame of the supernova, they are presented here in the observed frame. The fits did not include the last three epochs because of uncertainties in correcting for the flux of the host galaxy.}
\label{fig:ModelLCs}
\end{figure}

\subsection{Radioactive Decay Model}
\label{subsec:rad}

Semi-analytic modeling techniques for the powering of supernova light curves by the radioactive decay of $^{56}$Ni and $^{56}$Co have been in use for some time \citep{Arnett1982}. Here we implement a slightly modified form of the equations presented by \cite{CanoRadModel} where we account for possible gamma-ray leakage with the addition of a $( 1- e^{-A_{\gamma}t^{-2}} )$ factor. This factor is related to the gamma-ray optical depth and opacity of the ejecta via the expression $\tau_\gamma = \kappa_\gamma \rho R=A_{\gamma}t^{-2}$ \citep{ChatzopoulosCSMModel}. Larger values of $A_{\gamma}$ correspond to less gamma-ray leakage. We briefly consider radioactive decay models with complete gamma-ray trapping, but we find that such models are far more luminous (by a factor of $\sim 30$) at late-times than the observed light curve and do not consider them further.

The luminosity of the model is determined by
\begin{equation}
\tau_m = \sqrt[]{2} \bigg( \frac{\kappa}{\beta c} \frac{M_{ej}}{v_{ej}} \bigg)^{1/2},
\end{equation}
\begin{equation}
x=t/\tau_m, y=\tau_m/(2 \tau_{Ni}), s=\tau_m/(2 \tau_{Co}),
\end{equation}
\begin{equation}
A(z)=2ze^{-2zy+z^2}, B(z)=2ze^{-2zs+z^2}, \textrm{ and}
\end{equation}

\begin{multline}
L_{RD}(t)=M_{Ni}e^{-x^2} \bigg((\epsilon_{Ni}-\epsilon_{Co}) \int_{0}^{x} A(z)dz + \epsilon_{Co} \int_{0}^{x} B(z)dz \bigg) \\
\times \bigg( 1- e^{-A_{\gamma}(x \tau_m)^{-2}} \bigg),
\end{multline}
where $\beta \simeq 13.8$ is a constant of integration \citep{Arnett1982}, $c$ is the speed of light, $M_{ej}$ is the ejecta mass, $\kappa$ is the opacity of the ejecta, $v_{ej}$ is the ejecta velocity, and $\tau_m$ is the effective diffusion time. The decay times of $^{56}$Ni and $^{56}$Co are $\tau_{Ni}$=8.77 days and $\tau_{Co}$=111.3 days \citep{Taubenberger,Martin}, and the rates of energy release for $^{56}$Ni and $^{56}$Co are $\epsilon_{Ni}=3.90 \times 10^{10}$ ergs s$^{-1}$ g$^{-1}$ and $\epsilon_{Co}=6.78 \times 10^{9}$ ergs s$^{-1}$ g$^{-1}$ \citep{1984SutherlandWheeler,1997Cappellaro}. Like \cite{CanoRadModel} and \cite{KaramehmetogluModels}, we adopt an opacity of $\kappa=0.07$ cm$^2$ g$^{-1}$, which is a reasonable value for electron scattering in hydrogen-free ejecta \citep{Chugai2000,1992Chevalier}.

We also include a time offset parameter to account for the uncertainty in our explosion date estimates from Section~\ref{subsec:phot}. Since we cannot measure the ejecta velocity directly in our observed spectra due to the presence of the CSM, we adopt a value of $v_{ej}=7,000$ km/s, consistent with observations of ASASSN-15ed \citep{PastorelloVII}. By assuming this value for $v_{ej}$, we reduce the complexity of this model to four fitting parameters: $M_{ej}$, $M_{Ni}$, $A_{\gamma}$, and the explosion date. A best fit is then obtained by calculating a grid of models in this four dimensional parameter space and finding the model with the lowest ${\chi}^2$ value.

The result is shown by the gray curve in Figure~\ref{fig:ModelLCs}. The physical parameters are $M_{ej}=0.85 M_\odot$, $M_{Ni}=25.7 M_\odot$, and $A_{\gamma}=1.66\times 10^{12}$ s$^2$. The inferred explosion date of JD 2457015.3 agrees poorly with our crude estimate of JD $2456998.0$ in Section~\ref{subsec:phot}. Of the physical models we explored, this pure radioactive decay model produces the worst fit. Furthermore, the only way this model can reproduce the observed high peak luminosity is with $M_{Ni}>M_{ej}$, an unphysical condition. We also considered opacities up to 0.2 cm$^2$ g$^{-1}$ and values of $v_{ej}$ up to 20,000 km/s but find that such models still cannot produce good fits with physically reasonable parameters. Thus, although we know that some $^{56}$Ni must have been produced in the supernova explosion, its radioactive decay can be confidently ruled out as the primary power source for the light curve.

\subsection{Magnetar Spin Down Model}
\label{subsec:mag}

In this work we implement the magnetar spin down model of \cite{InserraMagModel}. We use simplified forms of the equations therein that have been informed by those presented by \cite{ChatzopoulosCSMModel}. Under the simplifying assumption of uniform ejecta density $E_k=3 M_{ej} v_{ej}^2/10$, where $E_k$ is the supernova's kinetic energy, $M_{ej}$ is the ejecta mass, and $v_{ej}$ is the ejecta velocity. The luminosity of the model is then described by
\begin{equation}
\tau_m = (6/5)^{1/4} \bigg(\frac{\kappa}{\beta c}\bigg)^{1/2} M_{ej}^{3/4} E_k^{-1/4},~\tau_p=4.7B_{14}^{-2}P_{ms}^2,
\end{equation}

\begin{equation}
x=t/\tau_m,~y=\tau_m/\tau_p, \textrm{ and}
\end{equation}

\begin{multline}
L_{Mag}(t)=\big(4.9\times10^{46} \textrm{ ergs s}^{-1} \big) B_{14}^2 P_{ms}^{-4} e^{-x^2} \\
\times \bigg( \int_{0}^{x} \frac{2ze^{z^2}}{(1+yz)^2}dz \bigg).
\end{multline}
The variables $\tau_m$, $\beta$, and $\kappa$ are the same as in the radioactive decay model, and we again adopt $\kappa=0.07$ cm$^2$ g$^{-1}$. The magnetar is described by its characteristic spin down timescale, $\tau_p$, the strength of its magnetic field, $B_{14}$, in units of $10^{14}$ G, and its initial rotational period, $P_{ms}$, in units of ms. We again include an explosion date offset parameter in our implementation of this model.

While the model fits improve with shorter initial rotation periods, we choose to limit our models to $P_{ms}=1.0$ ms, in order to be consistent with the theoretical limits on magnetar periods calculated by \cite{MetzgerSpinPeriod}. For this model we fit $v_{ej}$ as a fitting parameter with values constrained to be between 5,000 and 20,000 km/s, leading to a model with four fitting parameters: $M_{ej}$, $v_{ej}$, $B_{14}$, and the explosion date. We again calculate a grid of models in this four dimensional parameter space and obtain a best fit by selecting the model with the lowest ${\chi}^2$ value.

The result for this model is shown by the blue curve in Figure~\ref{fig:ModelLCs}. The physical parameters are $M_{ej}=3.29 M_\odot$, $v_{ej}=14,000$ km/s, $B = 17.2 \times 10^{14}$ G, and the imposed $P=1.0$ ms value. The model reproduces the observed light curve fairly well, and unlike the pure radioactive decay model, none of the parameters are obviously unphysical.

Although this model produces a fit in reasonable agreement with the observed light curve, there is little other evidence to support the hypothesis that ASASSN-14ms is powered by a magnetar. Additionally, the inferred explosion date of JD 2457017.9 is in poor agreement with our crude estimate of JD $2456998.0$ in Section~\ref{subsec:phot}. Furthermore, although it is not completely unphysical, a 1 ms initial spin period approaches theoretical limits, and models with considerably slower initial rotation do not match the bolometric light curve as well. As such, while we cannot rule out this model, we will focus instead on considering the contribution of CSM interaction, for which we have compelling spectroscopic evidence.

\subsection{Circumstellar Material Interaction Model}
\label{subsec:CSM}

We use the CSM-interaction model of \cite{CWV12} and \cite{ChatzopoulosCSMModel}, which is based on the work of \cite{1982Chevalier} and \cite{ChevalierFransson1994}. This particular formulation is a hybrid model combining radioactive decay plus circumstellar interactions (hence the two integrals in the expression for $L_{CSM}$ below), but it can trivially be reduced to a purely CSM-interaction model by setting $M_{Ni}=0$.

This model utilizes a CSM density profile of the form $\rho_{CSM}=q r^{-s}$. The exponent $s$ of the power-law is commonly set either to 0 (constant density CSM) or 2 (steady-wind CSM). The scaling factor $q$ is defined by $q=\rho_{CSM,1} R_p^s$, where $R_p$ is the radius of the progenitor star and $\rho_{CSM,1}$ is the density of the circumstellar medium immediately beyond this stellar radius. 

This model assigns a double power-law density profile to the supernova ejecta of the form $\rho_{SN}=g^n t^{n-3} r^{-n}$. The power-law exponent of the steeper outer component of the supernova ejecta is given by $n$ (with values of 6 to 14 being typical, as used by \cite{1982Chevalier}), while that of the inner component is given by $\delta$ (and is typically set to either 0 or 2). The scaling factor $g^n$, the total radius of the CSM, $R_{CSM}$, and the fixed photospheric radius of the CSM, $R_{ph}$, are given by
\begin{equation}
g^n = \frac{1}{4\pi (\delta-n)} \frac{(2(5-\delta)(n-5)E_{SN})^{(n-3)/2}}{((3-\delta)(n-3)M_{ej})^{(n-5)/2}},
\end{equation}

\begin{equation}
R_{CSM}=\bigg( \frac{M_{CSM}(3-s)}{4\pi q} + R_{p}^{3-s} \bigg)^{1/(3-s)}, \textrm{ and}
\end{equation}

\begin{equation}
R_{ph}=\bigg( -\frac{2(1-s)}{3\kappa q} +R_{CSM}^{1-s} \bigg)^{1/(1-s)}.
\end{equation}
Here, $E_{SN}$ denotes the total supernova energy, $M_{CSM}$ is the mass of circumstellar material, and $M_{ej}$ is the ejecta mass. The values of the constants $A$ (which differs from the gamma-ray leakage parameter $A_{\gamma}$), $\beta_F$, and $\beta_R$ are dependent on the values of $n$ and $s$ and can be found in Table 1 of \cite{1982Chevalier}, where $\beta_F=R_1/R_c$ and $\beta_R=R_2/R_c$. We adopt the assumptions used by \cite{KaramehmetogluModels} when modeling the SN Ibn OGLE-2014-SN-131 and take $s=0$ and $n=12$. The values of $A$, $\beta_F$, and $\beta_R$ are then $A=0.19$, $\beta_F=1.121$, and $\beta_R=0.974$.

The luminosity of this model is then set by
\begin{equation}
F(t') = \frac{2\pi}{(n-s)^3} g^{n \frac{5-s}{n-s}} q^\frac{n-5}{n-s} (n-3)^2 (n-5) \\
\beta_F^{5-s} A^\frac{5-s}{n-s} (t'+t_i)^\gamma,
\end{equation}

\begin{equation}
R(t') = 2\pi \bigg(\frac{Ag^n}{q}\bigg)^\frac{5-n}{n-s} \beta_R^{5-n}g^n\\
\bigg(\frac{3-s}{n-s}\bigg)^3 (t'-t_i)^\gamma, \textrm{ and}
\end{equation}

\begin{multline}
L_{CSM}(t)=\frac{M_{Ni}}{t'_0} e^{\frac{-t}{t'_0}} \int_{0}^{t} e^{\frac{t'}{t'_0}} \bigg[(\epsilon_{Ni}-\epsilon_{Co})e^{\frac{-t'}{t_{Ni}}}+\epsilon_{Co}e^{\frac{-t'}{t_{Co}}} \bigg] dt' \\
\times \bigg( 1- e^{-A_{\gamma}t^{-2}} \bigg) \\
+ \frac{e^{\frac{-t}{t_0}}}{t_0} \int_{0}^{t} e^{\frac{t'}{t_0}} \bigg[ F(t') \cdot \theta(t_{FS,BO}-t') + R(t') \cdot \theta(t_{RS,*}-t') \bigg] dt'.
\end{multline}
The first integral in this expression corresponds to the pure radioactive decay model described in Section~\ref{subsec:rad} as $L_{RD}(t)$, including the gamma-ray leakage term. The time of initial interaction between the CSM and the supernova ejecta is $t_i=R_p / v_{SN}$, where the characteristic velocity of the supernova ejecta is $v_{SN}=[10(n-5)E_{SN}/(3(n-3)M_{ej})]^{1/2}/x_0$. Here, $x_0$ is the dimensionless radius where the supernova ejecta density profile transitions from the inner profile ($\delta=2$) to the outer profile ($n=12$), and we use $x_0=0.7$. $\theta(t_{FS,BO}-t')$ and $\theta(t_{RS,*}-t')$ are Heaviside theta functions, and $t_{FS,*}$ and $t_{RS,*}$ are the forward shock and reverse shock termination timescales, respectively. We adopt the assumption of \cite{CWV12} that $t_{FS,BO}$ can be reasonably approximated by $t_{FS,*}$ when the optically thick component of the CSM mass ($M_{CSM,th}$) is used in the calculation. The shock termination times are then given by
\begin{equation}
M_{CSM,th} = \frac{4\pi q}{3-s} \big( R_{ph}^{3-s}- R_p^{3-s} \big),
\end{equation}

\begin{equation}
\xi = \frac{n-s}{(n-3)(3-s)},
\end{equation}

\begin{equation}
t_{FS,BO} \approx t_{FS,*} =\bigg| \frac{3-s}{4\pi\beta_F^{3-s}} q^\frac{3-n}{n-s} (Ag^n)^\frac{s-3}{n-s} \bigg|^\xi M_{CSM,th}^\xi, \textrm{ and}
\end{equation}

\begin{equation}
t_{RS,*}=\bigg| \frac{v_{SN}}{\beta_R(Ag^n/q)^{1/{(n-s)}}} \bigg( 1-\frac{(3-n)M_{ej}}{4\pi v_{SN}^{3-n}g^n} \bigg)^{1/(3-n)} \bigg|^\frac{n-s}{s-3}.
\end{equation}
As in the other models, $\beta \simeq 13.8$ is a constant of integration from \cite{Arnett1982}, $c$ is the speed of light, and $\kappa$ is the opacity. We adopt the assumption of \cite{KaramehmetogluModels} that the opacity is 0.20 cm$^2$ g$^{-1}$. The diffusion timescale expressions used in the pure radioactive decay and magnetar spin down models are valid only for the supernova ejecta. In this model we adopt more general expressions, necessary for considering diffusion through the CSM. The diffusion timescale through the optically thick component of the CSM, and that through the sum of the ejecta mass and the optically thick CSM are given by
\begin{equation}
t_0 = \frac{\kappa M_{CSM,th}}{\beta c R_{ph}} \textrm{  and  } t'_0 = \frac{\kappa (M_{ej}+M_{CSM,th})}{\beta c R_{ph}},
\end{equation}
respectively. This model is considerably more complex than the others. As such, we do not attempt to calculate a true best fit and instead simply present an optimized manual fit. We varied the numerous model parameters until we obtained a fit that reproduces the observed light curve reasonably well with physically justifiable input parameters. Excluding the three density profile exponents, there are seven fitting parameters: $M_{ej}$, $M_{CSM}$, $M_{Ni}$, $A_{\gamma}$, $E_{SN}$, $\rho_{CSM,1}$, and the explosion date. We generate a grid of models in this seven dimensional parameter space around the values used to produce the initial manual fit and then select the model with the lowest ${\chi}^2$ value.

This fit is shown by the gray curve in Figure~\ref{fig:ModelLCs}. The inferred explosion date of JD $2457008.5$ is significantly earlier than those of the other two models and better approximates our crude estimate of JD $2456998.0$ in Section~\ref{subsec:phot}. The physical parameters are $M_{ej}=4.28 M_\odot$, $M_{CSM}=0.51 M_\odot$, $M_{Ni}=0.23 M_\odot$, $A_{\gamma}=1.97 \times 10^{13}$ s$^2$, $E_{SN}=8.75 \times 10^{50}$ ergs, and $\rho_{CSM,1}=3.18 \times 10^{-13}$ g/cm$^3$. Such a large $M_{Ni}$ would be unusual for a supernova with $E_{SN}<10^{51}$ ergs, but it is not dramatically unphysical. We note that the photospheric radius of this fit ($\sim 10^{15}$ cm) is consistent with the early-phase radius estimates from our bolometric light curve (out to $\sim 40$ days beyond maximum light) and is broadly comparable to $R_{ph}$ values found using this physical model to fit other highly luminous SNe \citep{ChatzopoulosCSMModel}.

This model is able to reproduce the observed light curve extremely well across all epochs, although this may not be surprising given the number of fitting parameters. Still, the considerable spectral evidence for CSM interaction provides ample justification for its use. Considering both the spectral evidence and the quality fit of this physically reasonable model, we conclude that the light curve of ASASSN-14ms can be plausibly explained by a combination of CSM interaction (which dominates at early times) and radioactive decay (which dominates at late times).

Since this model reproduces the observed light curve well, we used it to make an estimate of the total radiated energy. Integrating this model over the rest-frame time period for which we have photometric observations yields a total radiated energy of $2.41 \times 10^{50}$ ergs, which is slightly larger than our more direct estimate of $(2.1 \pm 0.3) \times 10^{50}$ ergs. We can then expand this integral to calculate the total energy radiated from the time of initial explosion to 500 days later (when the supernova has faded to a negligible luminosity). Doing so, we obtain $2.66 \times 10^{50}$ ergs.

\section{Discussion and Conclusions}
\label{sec:conclusion}

We discuss the discovery of the highly luminous type Ibn supernova ASASSN-14ms and present follow-up observations spanning over one hundred days. This includes optical and UV photometry as well as 6 epochs of spectroscopy. We have also presented a spectrum of the host galaxy SDSS J130408.52+521846.4, which previously had no publicly available spectroscopic observations.

Spectroscopically, ASASSN-14ms is a broadly normal member of the Type Ibn supernova class. Its spectral features differ slightly from those of other SNe~Ibn at early times, but the class as a whole is fairly heterogeneous at this stage. By late times, ASASSN-14ms settles into the typical spectroscopic behavior of a SN Ibn. However, its light curve is broader and considerably more luminous than that of a typical SNe~Ibn, prompting us to investigate potential explanations for this peculiar feature.

After fitting three semi-analytic light curve models to the observed bolometric light curve, we conclude that it cannot be powered purely by the radioactive decay of $^{56}$Ni. Such models require an unphysically large nickel mass $(M_{Ni}>M_{ej})$. We conclude that a light curve powered purely by magnetar spin down is unlikely because such models require initial spin periods that are borderline unphysical and because we do not detect any observational signatures that indicate the formation of a magnetar. We note that the observed late-time behavior could likely be fit comparably well by magnetar spin down, but in the absence of other evidence for such a remnant we elected not to include this model.

A model light curve powered primarily by CSM interaction that includes the additional luminosity produced by the radioactive decay of a moderate amount of nickel ($\sim 0.25 M_\odot$) can reproduce the observed light curve of ASASSN-14ms quite well with physically reasonable input parameters. Although we cannot rule out a light curve powered by magnetar spin down, the CSM interaction model is well-motivated by the presence of intermediate-width He~I emission features in the spectra of SNe~Ibn (ASASSN-14ms included). Thus, we conclude that, although ASASSN-14ms is an unusually luminous member of the SNe~Ibn class, it can plausibly be explained by CSM interaction and radioactive decay with no need to invoke more exotic sources of energy deposition.

\section*{Acknowledgments}

We thank the referee for helpful comments. We thank the Las Cumbres Observatory and its staff for its continuing support of the ASAS-SN project. We also thank Swift PI, the Observation Duty Scientists, and the science planners for promptly approving and executing our observations.

ASAS-SN is supported by the Gordon and Betty Moore Foundation through grant
GBMF5490 to the Ohio State University and NSF grant AST-1515927. Development of
ASAS-SN has been supported by NSF grant AST-0908816, the Mt. Cuba Astronomical
Foundation, the Center for Cosmology and AstroParticle Physics at the Ohio
State University, the Chinese Academy of Sciences South America Center for
Astronomy (CAS- SACA), the Villum Foundation, and George Skestos.

The Liverpool Telescope is operated on the island of La Palma by Liverpool John Moores University in the Spanish Observatorio del Roque
de los Muchachos of the Instituto de Astrofisica de Canarias with financial
support from the UK Science and Technology Facilities Council. This research uses data products produced by the OIR Telescope Data Center, supported by the Smithsonian Astrophysical Observatory.

This research uses data obtained through the Telescope Access Program (TAP), which has been funded by ``the Strategic Priority Research Program-The Emergence of Cosmological Structures'' of the Chinese Academy of Sciences (Grant No.11 XDB09000000) and the Special Fund for Astronomy from the Ministry of Finance. This research was made possible through the use of the AAVSO Photometric All-Sky Survey (APASS), funded by the Robert Martin Ayers Sciences Fund.
 
Support for JLP is provided in part by FONDECYT through the grant 1151445 and by the Ministry of Economy, Development, and Tourism's Millennium Science Initiative through grant IC120009, awarded to The Millennium Institute of Astrophysics, MAS. KZS, CSK, and TAT are supported by NSF grants AST-1515876 and AST-1515927. TS is partly supported by NSF PHY-1404311 to John Beacom. PC and SD acknowledge Project 11573003 supported by NSFC. M. Stritzinger is supported by a research grant (13261) from VIL LUM FONDEN.


\begin{thebibliography}{70}
\expandafter\ifx\csname natexlab\endcsname\relax\def\natexlab#1{#1}\fi

\bibitem[{{Alam} {et~al}\mbox{.}(2015){Alam}, {Albareti}, {Allende Prieto},
  {Anders}, {Anderson}, {Anderton}, {Andrews}, {Armengaud}, {Aubourg},
  {Bailey}, \& et~al.}]{Alam15}
{Alam} S. {et~al.}, 2015, \apjs, 219, 12

\bibitem[{{Alard}(2000)}]{2000Alard}
{Alard} C., 2000, \aaps, 144, 363

\bibitem[{{Alard} \& {Lupton}(1998)}]{1998Alard}
{Alard} C., {Lupton} R.~H., 1998, \apj, 503, 325

\bibitem[{{Anupama} {et~al}\mbox{.}(2009){Anupama}, {Sahu}, {Gurugubelli},
  {Prabhu}, {Tominaga}, {Tanaka}, \& {Nomoto}}]{Anupama2006jc}
{Anupama} G.~C., {Sahu} D.~K., {Gurugubelli} U.~K., {Prabhu} T.~P., {Tominaga}
  N., {Tanaka} M., {Nomoto} K., 2009, \mnras, 392, 894

\bibitem[{{Arnett}(1982)}]{Arnett1982}
{Arnett} W.~D., 1982, \apj, 253, 785

\bibitem[{{Becker}, {White} \& {Helfand}(1995){Becker}, {White}, \&
  {Helfand}}]{Becker95}
{Becker} R.~H., {White} R.~L., {Helfand} D.~J., 1995, \apj, 450, 559

\bibitem[{{Blondin} \& {Tonry}(2007)}]{BlondinSNID}
{Blondin} S., {Tonry} J.~L., 2007, \apj, 666, 1024

\bibitem[{{Breeveld} {et~al}\mbox{.}(2011){Breeveld}, {Landsman}, {Holland},
  {Roming}, {Kuin}, \& {Page}}]{2011Breeveld}
{Breeveld} A.~A., {Landsman} W., {Holland} S.~T., {Roming} P., {Kuin} N.~P.~M.,
  {Page} M.~J., 2011, in American Institute of Physics Conference Series, Vol.
  1358, American Institute of Physics Conference Series, {McEnery} J.~E.,
  {Racusin} J.~L., {Gehrels} N., eds., pp. 373--376

\bibitem[{{Brown} {et~al}\mbox{.}(2013){Brown}, {Baliber}, {Bianco}, {Bowman},
  {Burleson}, {Conway}, {Crellin}, {Depagne}, {De Vera}, {Dilday}, {Dragomir},
  {Dubberley}, {Eastman}, {Elphick}, {Falarski}, {Foale}, {Ford}, {Fulton},
  {Garza}, {Gomez}, {Graham}, {Greene}, {Haldeman}, {Hawkins}, {Haworth},
  {Haynes}, {Hidas}, {Hjelstrom}, {Howell}, {Hygelund}, {Lister}, {Lobdill},
  {Martinez}, {Mullins}, {Norbury}, {Parrent}, {Paulson}, {Petry}, {Pickles},
  {Posner}, {Rosing}, {Ross}, {Sand}, {Saunders}, {Shobbrook}, {Shporer},
  {Street}, {Thomas}, {Tsapras}, {Tufts}, {Valenti}, {Vander Horst}, {Walker},
  {White}, \& {Willis}}]{Brown2013}
{Brown} T.~M. {et~al.}, 2013, \pasp, 125, 1031

\bibitem[{{Bruzual} \& {Charlot}(2003)}]{Bruzual03}
{Bruzual} G., {Charlot} S., 2003, \mnras, 344, 1000

\bibitem[{{Cano}(2013)}]{CanoRadModel}
{Cano} Z., 2013, \mnras, 434, 1098

\bibitem[{{Cappellaro} {et~al}\mbox{.}(1997){Cappellaro}, {Mazzali}, {Benetti},
  {Danziger}, {Turatto}, {della Valle}, \& {Patat}}]{1997Cappellaro}
{Cappellaro} E., {Mazzali} P.~A., {Benetti} S., {Danziger} I.~J., {Turatto} M.,
  {della Valle} M., {Patat} F., 1997, \aap, 328, 203

\bibitem[{{Cardelli}, {Clayton} \& {Mathis}(1989){Cardelli}, {Clayton}, \&
  {Mathis}}]{Cardelli89}
{Cardelli} J.~A., {Clayton} G.~C., {Mathis} J.~S., 1989, \apj, 345, 245

\bibitem[{{Chatzopoulos}, {Wheeler} \& {Vinko}(2012){Chatzopoulos}, {Wheeler},
  \& {Vinko}}]{CWV12}
{Chatzopoulos} E., {Wheeler} J.~C., {Vinko} J., 2012, \apj, 746, 121

\bibitem[{{Chatzopoulos} {et~al}\mbox{.}(2013){Chatzopoulos}, {Wheeler},
  {Vinko}, {Horvath}, \& {Nagy}}]{ChatzopoulosCSMModel}
{Chatzopoulos} E., {Wheeler} J.~C., {Vinko} J., {Horvath} Z.~L., {Nagy} A.,
  2013, \apj, 773, 76

\bibitem[{{Chevalier}(1982)}]{1982Chevalier}
{Chevalier} R.~A., 1982, \apj, 258, 790

\bibitem[{{Chevalier}(1992)}]{1992Chevalier}
---, 1992, \apj, 394, 599

\bibitem[{{Chevalier} \& {Fransson}(1994)}]{ChevalierFransson1994}
{Chevalier} R.~A., {Fransson} C., 1994, \apj, 420, 268

\bibitem[{{Chugai}(2000)}]{Chugai2000}
{Chugai} N.~N., 2000, Astronomy Letters, 26, 797

\bibitem[{{Condon} {et~al}\mbox{.}(1998){Condon}, {Cotton}, {Greisen}, {Yin},
  {Perley}, {Taylor}, \& {Broderick}}]{Condon98}
{Condon} J.~J., {Cotton} W.~D., {Greisen} E.~W., {Yin} Q.~F., {Perley} R.~A.,
  {Taylor} G.~B., {Broderick} J.~J., 1998, \aj, 115, 1693

\bibitem[{{Fabricant} {et~al}\mbox{.}(1998){Fabricant}, {Cheimets}, {Caldwell},
  \& {Geary}}]{FabricantFAST}
{Fabricant} D., {Cheimets} P., {Caldwell} N., {Geary} J., 1998, \pasp, 110, 79

\bibitem[{{Filippenko}(1997)}]{SNtypespaper}
{Filippenko} A.~V., 1997, \araa, 35, 309

\bibitem[{{Flewelling} {et~al}\mbox{.}(2016){Flewelling}, {Magnier},
  {Chambers}, {Heasley}, {Holmberg}, {Huber}, {Sweeney}, {Waters}, {Chen},
  {Farrow}, {Hasinger}, {Henderson}, {Long}, {Metcalfe}, {Nieto-Santisteban},
  {Norberg}, {Saglia}, {Szalay}, {Rest}, {Thakar}, {Tonry}, {Valenti},
  {Werner}, {White}, {Denneau}, {Draper}, {Hodapp}, {Jedicke}, {Kaiser},
  {Kudritzki}, {Price}, {Wainscoat}, {Chastel}, {McClean}, {Postman}, \&
  {Shiao}}]{PanSTARRS2016Flewelling}
{Flewelling} H.~A. {et~al.}, 2016, ArXiv e-prints

\bibitem[{{Foley} {et~al}\mbox{.}(2007){Foley}, {Smith}, {Ganeshalingam}, {Li},
  {Chornock}, \& {Filippenko}}]{Foley2006jc}
{Foley} R.~J., {Smith} N., {Ganeshalingam} M., {Li} W., {Chornock} R.,
  {Filippenko} A.~V., 2007, \apjl, 657, L105

\bibitem[{{Godoy-Rivera} {et~al}\mbox{.}(2017){Godoy-Rivera}, {Stanek},
  {Kochanek}, {Chen}, {Dong}, {Prieto}, {Shappee}, {Jha}, {Foley}, {Pan},
  {Holoien}, {Thompson}, {Grupe}, \& {Beacom}}]{Godoy2017}
{Godoy-Rivera} D. {et~al.}, 2017, \mnras, 466, 1428

\bibitem[{{Gorbikov} {et~al}\mbox{.}(2014){Gorbikov}, {Gal-Yam}, {Ofek},
  {Vreeswijk}, {Nugent}, {Chotard}, {Kulkarni}, {Cao}, {De Cia}, {Yaron},
  {Tal}, {Arcavi}, {Kasliwal}, {Cenko}, {Sullivan}, \&
  {Chen}}]{iPTF13beoGorbikov}
{Gorbikov} E. {et~al.}, 2014, \mnras, 443, 671

\bibitem[{{Henden} {et~al}\mbox{.}(2015){Henden}, {Levine}, {Terrell}, \&
  {Welch}}]{Henden2015}
{Henden} A.~A., {Levine} S., {Terrell} D., {Welch} D.~L., 2015, in American
  Astronomical Society Meeting Abstracts, Vol. 225, American Astronomical
  Society Meeting Abstracts, p. 336.16

\bibitem[{{Holoien} {et~al}\mbox{.}(2017){Holoien}, {Stanek}, {Kochanek},
  {Shappee}, {Prieto}, {Brimacombe}, {Bersier}, {Bishop}, {Dong}, {Brown},
  {Danilet}, {Simonian}, {Basu}, {Beacom}, {Falco}, {Pojmanski}, {Skowron},
  {Wo{\'z}niak}, {{\'A}vila}, {Conseil}, {Contreras}, {Cruz}, {Fern{\'a}ndez},
  {Koff}, {Guo}, {Herczeg}, {Hissong}, {Hsiao}, {Jose}, {Kiyota}, {Long},
  {Monard}, {Nicholls}, {Nicolas}, \& {Wiethoff}}]{ASASSNCatalog}
{Holoien} T.~W.-S. {et~al.}, 2017, \mnras, 464, 2672

\bibitem[{{Hosseinzadeh} {et~al}\mbox{.}(2017){Hosseinzadeh}, {Arcavi},
  {Valenti}, {McCully}, {Howell}, {Johansson}, {Sollerman}, {Pastorello},
  {Benetti}, {Cao}, {Cenko}, {Clubb}, {Corsi}, {Duggan}, {Elias-Rosa},
  {Filippenko}, {Fox}, {Fremling}, {Horesh}, {Karamehmetoglu}, {Kasliwal},
  {Marion}, {Ofek}, {Sand}, {Taddia}, {Zheng}, {Fraser}, {Gal-Yam}, {Inserra},
  {Laher}, {Masci}, {Rebbapragada}, {Smartt}, {Smith}, {Sullivan}, {Surace}, \&
  {Wo{\'z}niak}}]{HosseinzadehIbnMaxLight}
{Hosseinzadeh} G. {et~al.}, 2017, \apj, 836, 158

\bibitem[{{Inserra} {et~al}\mbox{.}(2013){Inserra}, {Smartt}, {Jerkstrand},
  {Valenti}, {Fraser}, {Wright}, {Smith}, {Chen}, {Kotak}, {Pastorello},
  {Nicholl}, {Bresolin}, {Kudritzki}, {Benetti}, {Botticella}, {Burgett},
  {Chambers}, {Ergon}, {Flewelling}, {Fynbo}, {Geier}, {Hodapp}, {Howell},
  {Huber}, {Kaiser}, {Leloudas}, {Magill}, {Magnier}, {McCrum}, {Metcalfe},
  {Price}, {Rest}, {Sollerman}, {Sweeney}, {Taddia}, {Taubenberger}, {Tonry},
  {Wainscoat}, {Waters}, \& {Young}}]{InserraMagModel}
{Inserra} C. {et~al.}, 2013, \apj, 770, 128

\bibitem[{{Karamehmetoglu} {et~al}\mbox{.}(2017){Karamehmetoglu}, {Taddia},
  {Sollerman}, {Wyrzykowski}, {Schmidl}, {Fraser}, {Fremling}, {Greiner},
  {Inserra}, {Kostrzewa-Rutkowska}, {Maguire}, {Smartt}, {Sullivan}, \&
  {Young}}]{KaramehmetogluModels}
{Karamehmetoglu} E. {et~al.}, 2017, \aap, 602, A93

\bibitem[{{Kennicutt}(1998)}]{Kennicutt98}
{Kennicutt}, Jr. R.~C., 1998, \araa, 36, 189

\bibitem[{{Kiyota} {et~al}\mbox{.}(2014){Kiyota}, {Holoien}, {Stanek},
  {Kochanek}, {Davis}, {Simonian}, {Basu}, {Goss}, {Beacom}, {Shappee},
  {Prieto}, {Bersier}, {Dong}, {Wozniak}, {Brimacombe}, {Szczygiel},
  {Pojmanski}, {Conseil}, {Koff}, {Nicholls}, \& {Nicolas}}]{14msDiscoveryATel}
{Kiyota} S. {et~al.}, 2014, The Astronomer's Telegram, 6853

\bibitem[{{Kriek} {et~al}\mbox{.}(2009){Kriek}, {van Dokkum}, {Labb{\'e}},
  {Franx}, {Illingworth}, {Marchesini}, \& {Quadri}}]{Kriek09}
{Kriek} M., {van Dokkum} P.~G., {Labb{\'e}} I., {Franx} M., {Illingworth}
  G.~D., {Marchesini} D., {Quadri} R.~F., 2009, \apj, 700, 221

\bibitem[{{Martin} {et~al}\mbox{.}(2005){Martin}, {Fanson}, {Schiminovich},
  {Morrissey}, {Friedman}, {Barlow}, {Conrow}, {Grange}, {Jelinsky},
  {Milliard}, {Siegmund}, {Bianchi}, {Byun}, {Donas}, {Forster}, {Heckman},
  {Lee}, {Madore}, {Malina}, {Neff}, {Rich}, {Small}, {Surber}, {Szalay},
  {Welsh}, \& {Wyder}}]{Martin05}
{Martin} D.~C. {et~al.}, 2005, \apjl, 619, L1

\bibitem[{{Martin}(1987)}]{Martin}
{Martin} M.~J.~e., 1987, 58, 67

\bibitem[{{Martini} {et~al}\mbox{.}(2011){Martini}, {Stoll}, {Derwent},
  {Zhelem}, {Atwood}, {Gonzalez}, {Mason}, {O'Brien}, {Pappalardo}, {Pogge},
  {Ward}, \& {Wong}}]{MartiniOSMOS}
{Martini} P. {et~al.}, 2011, \pasp, 123, 187

\bibitem[{{Matheson} {et~al}\mbox{.}(2000){Matheson}, {Filippenko}, {Chornock},
  {Leonard}, \& {Li}}]{99cqMatheson}
{Matheson} T., {Filippenko} A.~V., {Chornock} R., {Leonard} D.~C., {Li} W.,
  2000, \aj, 119, 2303

\bibitem[{{Metzger} {et~al}\mbox{.}(2015){Metzger}, {Margalit}, {Kasen}, \&
  {Quataert}}]{MetzgerSpinPeriod}
{Metzger} B.~D., {Margalit} B., {Kasen} D., {Quataert} E., 2015, \mnras, 454,
  3311

\bibitem[{{Moriya} \& {Maeda}(2016)}]{Moriya2010al}
{Moriya} T.~J., {Maeda} K., 2016, \apj, 824, 100

\bibitem[{{Nakano} {et~al}\mbox{.}(2006){Nakano}, {Itagaki}, {Puckett}, \&
  {Gorelli}}]{2006CBETNakano}
{Nakano} S., {Itagaki} K., {Puckett} T., {Gorelli} R., 2006, Central Bureau
  Electronic Telegrams, 666

\bibitem[{{Nugent} {et~al}\mbox{.}(2011){Nugent}, {Sullivan}, {Cenko},
  {Thomas}, {Kasen}, {Howell}, {Bersier}, {Bloom}, {Kulkarni}, {Kandrashoff},
  {Filippenko}, {Silverman}, {Marcy}, {Howard}, {Isaacson}, {Maguire},
  {Suzuki}, {Tarlton}, {Pan}, {Bildsten}, {Fulton}, {Parrent}, {Sand},
  {Podsiadlowski}, {Bianco}, {Dilday}, {Graham}, {Lyman}, {James}, {Kasliwal},
  {Law}, {Quimby}, {Hook}, {Walker}, {Mazzali}, {Pian}, {Ofek}, {Gal-Yam}, \&
  {Poznanski}}]{Nugent2011}
{Nugent} P.~E. {et~al.}, 2011, \nat, 480, 344

\bibitem[{{Pastorello} {et~al}\mbox{.}(2015{\natexlab{a}}){Pastorello},
  {Benetti}, {Brown}, {Tsvetkov}, {Inserra}, {Taubenberger}, {Tomasella},
  {Fraser}, {Rich}, {Botticella}, {Bufano}, {Cappellaro}, {Ergon},
  {Gorbovskoy}, {Harutyunyan}, {Huang}, {Kotak}, {Lipunov}, {Magill},
  {Miluzio}, {Morrell}, {Ochner}, {Smartt}, {Sollerman}, {Spiro},
  {Stritzinger}, {Turatto}, {Valenti}, {Wang}, {Wright}, {Yurkov}, {Zampieri},
  \& {Zhang}}]{PastorelloIV}
{Pastorello} A. {et~al.}, 2015{\natexlab{a}}, \mnras, 449, 1921

\bibitem[{{Pastorello} {et~al}\mbox{.}(2015{\natexlab{b}}){Pastorello},
  {Hadjiyska}, {Rabinowitz}, {Valenti}, {Turatto}, {Fasano}, {Benitez-Herrera},
  {Baltay}, {Benetti}, {Botticella}, {Cappellaro}, {Elias-Rosa}, {Ellman},
  {Feindt}, {Filippenko}, {Fraser}, {Gal-Yam}, {Graham}, {Howell}, {Inserra},
  {Kelly}, {Kotak}, {Kowalski}, {McKinnon}, {Morales-Garoffolo}, {Nugent},
  {Smartt}, {Smith}, {Stritzinger}, {Sullivan}, {Taubenberger}, {Walker},
  {Yaron}, \& {Young}}]{PastorelloVI}
---, 2015{\natexlab{b}}, \mnras, 449, 1954

\bibitem[{{Pastorello} {et~al}\mbox{.}(2008{\natexlab{a}}){Pastorello},
  {Mattila}, {Zampieri}, {Della Valle}, {Smartt}, {Valenti}, {Agnoletto},
  {Benetti}, {Benn}, {Branch}, {Cappellaro}, {Dennefeld}, {Eldridge},
  {Gal-Yam}, {Harutyunyan}, {Hunter}, {Kjeldsen}, {Lipkin}, {Mazzali}, {Milne},
  {Navasardyan}, {Ofek}, {Pian}, {Shemmer}, {Spiro}, {Stathakis},
  {Taubenberger}, {Turatto}, \& {Yamaoka}}]{PastorelloIbn}
---, 2008{\natexlab{a}}, \mnras, 389, 113

\bibitem[{{Pastorello} {et~al}\mbox{.}(2015{\natexlab{c}}){Pastorello},
  {Prieto}, {Elias-Rosa}, {Bersier}, {Hosseinzadeh}, {Morales-Garoffolo},
  {Noebauer}, {Taubenberger}, {Tomasella}, {Kochanek}, {Falco}, {Basu},
  {Beacom}, {Benetti}, {Brimacombe}, {Cappellaro}, {Danilet}, {Dong},
  {Fernandez}, {Goss}, {Granata}, {Harutyunyan}, {Holoien}, {Ishida}, {Kiyota},
  {Krannich}, {Nicholls}, {Ochner}, {Pojma{\'n}ski}, {Shappee}, {Simonian},
  {Stanek}, {Starrfield}, {Szczygie{\l}}, {Tartaglia}, {Terreran}, {Thompson},
  {Turatto}, {Wagner}, {Wiethoff}, {Wilber}, \& {Wo{\'z}niak}}]{PastorelloVII}
---, 2015{\natexlab{c}}, \mnras, 453, 3649

\bibitem[{{Pastorello} {et~al}\mbox{.}(2008{\natexlab{b}}){Pastorello},
  {Quimby}, {Smartt}, {Mattila}, {Navasardyan}, {Crockett}, {Elias-Rosa},
  {Mondol}, {Wheeler}, \& {Young}}]{Pastorello2005la}
---, 2008{\natexlab{b}}, \mnras, 389, 131

\bibitem[{{Pastorello} {et~al}\mbox{.}(2007){Pastorello}, {Smartt}, {Mattila},
  {Eldridge}, {Young}, {Itagaki}, {Yamaoka}, {Navasardyan}, {Valenti}, {Patat},
  {Agnoletto}, {Augusteijn}, {Benetti}, {Cappellaro}, {Boles}, {Bonnet-Bidaud},
  {Botticella}, {Bufano}, {Cao}, {Deng}, {Dennefeld}, {Elias-Rosa},
  {Harutyunyan}, {Keenan}, {Iijima}, {Lorenzi}, {Mazzali}, {Meng}, {Nakano},
  {Nielsen}, {Smoker}, {Stanishev}, {Turatto}, {Xu}, \&
  {Zampieri}}]{Pastorello06jcOutburst}
---, 2007, \nat, 447, 829

\bibitem[{{Pastorello} {et~al}\mbox{.}(2015{\natexlab{d}}){Pastorello},
  {Tartaglia}, {Elias-Rosa}, {Morales-Garoffolo}, {Terreran}, {Taubenberger},
  {Noebauer}, {Benetti}, {Cappellaro}, {Ciabattari}, {Dennefeld}, {Dimai},
  {Ishida}, {Harutyunyan}, {Leonini}, {Ochner}, {Sollerman}, {Taddia}, \&
  {Zaggia}}]{PastorelloVIII}
---, 2015{\natexlab{d}}, \mnras, 454, 4293

\bibitem[{{Pastorello} {et~al}\mbox{.}(2015{\natexlab{e}}){Pastorello},
  {Wyrzykowski}, {Valenti}, {Prieto}, {Koz{\l}owski}, {Udalski}, {Elias-Rosa},
  {Morales-Garoffolo}, {Anderson}, {Benetti}, {Bersten}, {Botticella},
  {Cappellaro}, {Fasano}, {Fraser}, {Gal-Yam}, {Gillone}, {Graham}, {Greiner},
  {Hachinger}, {Howell}, {Inserra}, {Parrent}, {Rau}, {Schulze}, {Smartt},
  {Smith}, {Turatto}, {Yaron}, {Young}, {Kubiak}, {Szyma{\'n}ski},
  {Pietrzy{\'n}ski}, {Soszy{\'n}ski}, {Ulaczyk}, {Poleski}, {Pietrukowicz},
  {Skowron}, \& {Mr{\'o}z}}]{PastorelloV}
---, 2015{\natexlab{e}}, \mnras, 449, 1941

\bibitem[{{Pettini} \& {Pagel}(2004)}]{Pettini04}
{Pettini} M., {Pagel} B.~E.~J., 2004, \mnras, 348, L59

\bibitem[{{Pogge} {et~al}\mbox{.}(2010){Pogge}, {Atwood}, {Brewer}, {Byard},
  {Derwent}, {Gonzalez}, {Martini}, {Mason}, {O'Brien}, {Osmer}, {Pappalardo},
  {Steinbrecher}, {Teiga}, \& {Zhelem}}]{PoggeMODS}
{Pogge} R.~W. {et~al.}, 2010, in \procspie, Vol. 7735, Ground-based and
  Airborne Instrumentation for Astronomy III, p. 77350A

\bibitem[{{Prieto} {et~al}\mbox{.}(2014){Prieto}, {Shappee}, {Holoien},
  {Stanek}, {Davis}, {Kochanek}, {Jencson}, {Basu}, {Beacom}, {Bersier},
  {Szczygiel}, {Pojmanski}, \& {Brimacombe}}]{14ddClassification}
{Prieto} J.~L. {et~al.}, 2014, The Astronomer's Telegram, 6293

\bibitem[{{Riess} {et~al}\mbox{.}(1999){Riess}, {Filippenko}, {Li}, {Treffers},
  {Schmidt}, {Qiu}, {Hu}, {Armstrong}, {Faranda}, {Thouvenot}, \&
  {Buil}}]{Riess1999}
{Riess} A.~G. {et~al.}, 1999, \aj, 118, 2675

\bibitem[{{Roming} {et~al}\mbox{.}(2005){Roming}, {Kennedy}, {Mason}, {Nousek},
  {Ahr}, {Bingham}, {Broos}, {Carter}, {Hancock}, {Huckle}, {Hunsberger},
  {Kawakami}, {Killough}, {Koch}, {McLelland}, {Smith}, {Smith}, {Soto},
  {Boyd}, {Breeveld}, {Holland}, {Ivanushkina}, {Pryzby}, {Still}, \&
  {Stock}}]{2005RomingUVOT}
{Roming} P.~W.~A. {et~al.}, 2005, \ssr, 120, 95

\bibitem[{{Schlafly} \& {Finkbeiner}(2011)}]{Schlafly11}
{Schlafly} E.~F., {Finkbeiner} D.~P., 2011, \apj, 737, 103

\bibitem[{{Schlegel}, {Finkbeiner} \& {Davis}(1998){Schlegel}, {Finkbeiner}, \&
  {Davis}}]{1998Schlegel}
{Schlegel} D.~J., {Finkbeiner} D.~P., {Davis} M., 1998, \apj, 500, 525

\bibitem[{{Schlegel}(1990)}]{TypeIInSchlegel}
{Schlegel} E.~M., 1990, \mnras, 244, 269

\bibitem[{{Shappee} {et~al}\mbox{.}(2014){Shappee}, {Prieto}, {Grupe},
  {Kochanek}, {Stanek}, {De Rosa}, {Mathur}, {Zu}, {Peterson}, {Pogge},
  {Komossa}, {Im}, {Jencson}, {Holoien}, {Basu}, {Beacom}, {Szczygie{\l}},
  {Brimacombe}, {Adams}, {Campillay}, {Choi}, {Contreras}, {Dietrich},
  {Dubberley}, {Elphick}, {Foale}, {Giustini}, {Gonzalez}, {Hawkins}, {Howell},
  {Hsiao}, {Koss}, {Leighly}, {Morrell}, {Mudd}, {Mullins}, {Nugent},
  {Parrent}, {Phillips}, {Pojmanski}, {Rosing}, {Ross}, {Sand}, {Terndrup},
  {Valenti}, {Walker}, \& {Yoon}}]{ShappeeASASSN}
{Shappee} B.~J. {et~al.}, 2014, \apj, 788, 48

\bibitem[{{Shivvers} {et~al}\mbox{.}(2016){Shivvers}, {Zheng}, {Mauerhan},
  {Kleiser}, {Van Dyk}, {Silverman}, {Graham}, {Kelly}, {Filippenko}, \&
  {Kumar}}]{2015UShivvers}
{Shivvers} I. {et~al.}, 2016, \mnras, 461, 3057

\bibitem[{{Smartt} {et~al}\mbox{.}(2016){Smartt}, {Chambers}, {Smith}, {Huber},
  {Young}, {Chen}, {Inserra}, {Wright}, {Coughlin}, {Denneau}, {Flewelling},
  {Heinze}, {Jerkstrand}, {Magnier}, {Maguire}, {Mueller}, {Rest}, {Sherstyuk},
  {Stalder}, {Schultz}, {Stubbs}, {Tonry}, {Waters}, {Wainscoat}, {Della
  Valle}, {Dennefeld}, {Dimitriadis}, {Firth}, {Fraser}, {Frohmaier},
  {Gal-Yam}, {Harmanen}, {Kankare}, {Kotak}, {Kromer}, {Mandel}, {Sollerman},
  {Gibson}, {Primak}, \& {Willman}}]{2016SmarttPS15dpn}
{Smartt} S.~J. {et~al.}, 2016, \apjl, 827, L40

\bibitem[{{Smith} {et~al}\mbox{.}(2012){Smith}, {Mauerhan}, {Silverman},
  {Ganeshalingam}, {Filippenko}, {Cenko}, {Clubb}, \&
  {Kandrashoff}}]{Smith2011hw}
{Smith} N., {Mauerhan} J.~C., {Silverman} J.~M., {Ganeshalingam} M.,
  {Filippenko} A.~V., {Cenko} S.~B., {Clubb} K.~I., {Kandrashoff} M.~T., 2012,
  \mnras, 426, 1905

\bibitem[{{Steele} {et~al}\mbox{.}(2004){Steele}, {Smith}, {Rees}, {Baker},
  {Bates}, {Bode}, {Bowman}, {Carter}, {Etherton}, {Ford}, {Fraser}, {Gomboc},
  {Lett}, {Mansfield}, {Marchant}, {Medrano-Cerda}, {Mottram}, {Raback},
  {Scott}, {Tomlinson}, \& {Zamanov}}]{Steele2004lT}
{Steele} I.~A. {et~al.}, 2004, in \procspie, Vol. 5489, Ground-based
  Telescopes, {Oschmann} Jr. J.~M., ed., pp. 679--692

\bibitem[{{Sutherland} \& {Wheeler}(1984)}]{1984SutherlandWheeler}
{Sutherland} P.~G., {Wheeler} J.~C., 1984, \apj, 280, 282

\bibitem[{{Taddia} {et~al}\mbox{.}(2015){Taddia}, {Sollerman}, {Fremling},
  {Pastorello}, {Leloudas}, {Fransson}, {Nyholm}, {Stritzinger}, {Ergon},
  {Roy}, \& {Migotto}}]{TaddiaMetallicity}
{Taddia} F. {et~al.}, 2015, \aap, 580, A131

\bibitem[{{Taubenberger} {et~al}\mbox{.}(2006){Taubenberger}, {Pastorello},
  {Mazzali}, {Valenti}, {Pignata}, {Sauer}, {Arbey}, {B{\"a}rnbantner},
  {Benetti}, {Della Valle}, {Deng}, {Elias-Rosa}, {Filippenko}, {Foley},
  {Goobar}, {Kotak}, {Li}, {Meikle}, {Mendez}, {Patat}, {Pian}, {Ries},
  {Ruiz-Lapuente}, {Salvo}, {Stanishev}, {Turatto}, \&
  {Hillebrandt}}]{Taubenberger}
{Taubenberger} S. {et~al.}, 2006, \mnras, 371, 1459

\bibitem[{{Tokarz} \& {Roll}(1997)}]{1997Tokarz}
{Tokarz} S.~P., {Roll} J., 1997, in Astronomical Society of the Pacific
  Conference Series, Vol. 125, Astronomical Data Analysis Software and Systems
  VI, {Hunt} G., {Payne} H., eds., p. 140

\bibitem[{{Tonry} \& {Davis}(1979)}]{TonrySNIDAlgorithm}
{Tonry} J., {Davis} M., 1979, \aj, 84, 1511

\bibitem[{{Tsvetkov}, {Volkov} \& {Pavlyuk}(2015){Tsvetkov}, {Volkov}, \&
  {Pavlyuk}}]{TsvetkovPSN}
{Tsvetkov} D.~Y., {Volkov} I.~M., {Pavlyuk} N.~N., 2015, Information Bulletin
  on Variable Stars, 6140

\bibitem[{{Wright}(2006)}]{2006Wright}
{Wright} E.~L., 2006, \pasp, 118, 1711

\end{thebibliography}

\label{lastpage}

\end{document}